\documentclass[twocolumn,floatfix]{aastex63}

\usepackage{amsmath,amssymb,mathtools}
\usepackage[textsize=small]{todonotes}
\usepackage{savesym}
\savesymbol{tablenum}
\usepackage{siunitx}
\restoresymbol{SIX}{tablenum}
\usepackage{subfigure}
\usepackage{enumitem}

\newcommand{\ee}{{\rm e}}

\newcommand{\nablab}{{\boldsymbol{\nabla}}}

\newcommand{\Ha}{{\rm Ha}}%
\newcommand{\Pe}{{\rm Pe}}%
\newcommand{\Ro}{{\rm Ro}}%
\newcommand{\Se}{{\rm Se}}%

\DeclareSIUnit{\gauss}{G}
\DeclareSIUnit{\year}{yr}
\DeclareSIUnit{\erg}{erg}

\received{\today}
\revised{\today}
\accepted{\today}
\submitjournal{ApJ}

\shorttitle{3D magneto-thermal evolution of neutron stars}
\shortauthors{De Grandis et al.}
\graphicspath{{./}{figures/}}
\begin{document}

\title{3D modelling of magneto-thermal evolution of neutron stars: method and test cases}

\correspondingauthor{Davide De Grandis}
\email{davide.degrandis@phd.unipd.it}

\author{Davide De Grandis}
\affiliation{Department of Physics and Astronomy, University of Padova, via Marzolo 8, 
I-35131 Padova, Italy}
\affiliation{Mullard Space Science Laboratory, University College London, 
Holmbury St. Mary, Surrey,
RH5 6NT, United Kingdom}

\author{Roberto Turolla}
\affiliation{Department of Physics and Astronomy, University of Padova, via Marzolo 8, 
I-35131 Padova, Italy}
\affiliation{Mullard Space Science Laboratory, University College London, 
Holmbury St. Mary, Surrey,
RH5 6NT, United Kingdom}

\author{Toby S. Wood}
\affiliation{School of Mathematics and Statistics, Newcastle University, 
Newcastle upon Tyne,
NE1 7RU, United Kingdom}

\author{Silvia Zane}
\affiliation{Mullard Space Science Laboratory, University College London,
Holmbury St. Mary, Surrey, RH5 6NT, United Kingdom}

\author{Roberto Taverna}
\affiliation{Department of Mathematics and Physics, University of Roma Tre,
via della Vasca Navale 84, I-00146 Roma, Italy}
\affiliation{Department of Physics and Astronomy, University of Padova, via Marzolo 8, 
I-35131 Padova, Italy}

\author{Konstantinos N. Gourgouliatos}
\affiliation{Department  of  Physics,  University  of  Patras,  Patras  26500,  Greece}
\affiliation{Department  of  Mathematical  Sciences,  Durham  University,  Durham  DH1 3LE, United Kingdom}

\begin{abstract}
Neutron stars harbour extremely strong magnetic fields within their solid outer crust. The topology of this field strongly influences the surface temperature distribution, and hence the star's observational properties. In this work, we present the first realistic simulations of the coupled crustal magneto-thermal evolution of isolated neutron stars in three dimensions with account for neutrino emission, obtained with the pseudo-spectral code {\sc parody}. We investigate both the secular evolution, especially in connection with the onset of instabilities during the Hall phase, and the short-term evolution following episodes of localised energy injection. Simulations show that a resistive tearing instability develops in about a Hall time if the initial toroidal field exceeds $\approx 10^{15}$ G. This leads to crustal failures because of the huge magnetic stresses coupled with the local temperature enhancement produced by dissipation. Localised heat deposition in the crust results in the appearance of hot spots on the star surface which can exhibit a variety of patterns. Since the transport properties are strongly influenced by the magnetic field, the hot regions tend to drift away and get deformed following the magnetic field lines while cooling. The shapes obtained with our simulations are reminiscent of those recently derived from NICER X-ray observations of the millisecond pulsar {PSR~J0030+0451}.
\end{abstract}


\section{Introduction}
Neutron stars (NSs) are unanimously believed to power some of the most violent phenomena observed in the high-energy sky, from the hyper-energetic giant flares of ultra-magnetised NSs \citep[\emph{magnetars}; see e.g.][for reviews]{2015RPPh...78k6901T, 2017ARA&A..55..261K}, to the spectacular merging of a binary NS system and the associated emission of gravitational waves \citep{2017ApJ...848L..13A}. Despite this, many aspects of NS physics are still poorly understood, mainly---but not only--- concerning their internal structure and composition, as well as the topology of their magnetic field. 

Isolated NSs, from which (thermal) emission coming directly from star surface is visible in the X-ray-to-optical bands, provide an ideal laboratory to investigate the physics of the interior of these objects, as first suggested by
\citet[see also \citealt{2009ASSL..357..141T}]{tsuruta}. NSs cool down as they age and their thermal evolution is coupled to that of their magnetic field. Knowledge of the secular magneto-thermal evolution can discriminate between different cooling scenarios when compared to observations, thus constraining the equation of state of ultra-dense matter \cite[see e.g.][]{2006NuPhA.777..497P, 2007ASSL..326.....H}. Moreover, it provides a self-consistent map of the surface temperature, which is essential in deriving any reliable estimate of the star radius from X-ray observations of (passively) cooling NSs \citep[see e.g.][and references therein]{2019MNRAS.485.5363G}.
A detailed model of the short-term evolution, following an impulsive energy release in the NS surface layers, is equally desirable since it directly bears to the origin of magnetar outbursts \cite[see e.g.][]{2011ASSP...21..247R, 2012ApJ...750L...6P,2018MNRAS.474..961C} and
of thermal X-ray emission in radio pulsars \cite[see e.g.][]{2009ASSL..357...91B,miller}.

Magneto-thermal evolution of NSs has been the focus of many investigations over the past decades \cite[see][for a complete historical outline and further references]{2013PhDT.........7V}. First attempts dealt with cooling in one dimensional (i.e., spherically symmetric) models with little or no account for the magnetic field \citep[see e.g.][]{1981SvAL....7...88Y, 1990ApJ...354L..17P}. As a further step, axysimmetic, 2D, calculations were produced, but these either assumed a known evolution of the temperature when solving for the magnetic field \citep{2007A&A...470..303P} or the opposite \citep{2008A&A...486..255A}. Moreover, inherent numerical difficulties prevented for a long time from including the Hall term in the induction equation, despite its importance in rearranging the magnetic field on the smaller spatial scales where dissipation is faster \citep{2007A&A...470..303P}. The first consistent treatment of the coupled magneto-thermal evolution in 2D was presented in \citet{2013MNRAS.434..123V}, who also succeeded in coping with the Hall term. Recent efforts were devoted to investigate the magnetic evolution with a fully 3D approach and confirmed the role of the Hall term in shaping the magnetic field in the earlier stages of the NS evolution when a peculiar magnetic structure develops \cite[the \emph{Hall attractor};][]{2016PNAS}.

According to the commonly accepted picture, the core of NSs is in a superfluid and superconducting state, for which the ground state is magnetic flux-free. \added{Up to now very few investigations dealt with the  magnetic evolution including the core \citep[see e.g.][]{2013MNRAS.435L..43C} and the structure of the magnetic field in the core of a NS is poorly understood as yet. Most studies of the magnetic field evolution, both in two and three dimensions, were restricted to the NS crust, relying on the assumption that the Meissner effect has been able to expel any flux from the core in a timescale shorter than those of magnetic and thermal evolution \citetext{see e.g. \citealt{2014MNRAS.437..424L}, but also \citealt{Ho_2017} for a different perspective}}. In this work the same approach is followed.

Over the last few years X-ray \citep[e.g.][]{2019ApJ...887L..23B} observations provided increasing evidence for the presence of localised region(s) on the surface of different classes of isolated NSs with non trivial thermal/magnetic properties and evolution. To explain these observations, as well as to validate results obtained in 2D calculations, fully coupled magneto-thermal 3D simulations are necessary. In this work we present some of the first simulations of such kind, showing some of the possible applications in which a 3D treatment is necessary to fully tackle the observed phenomenology.

The paper is organised as follows. In section \ref{sec:equations} we present the basic equations and their numerical implementation. In section \ref{sec:cases} some study cases for the long-term evolution of NSs are presented; in particular, the onset of eMHD instabilities is discussed in section \ref{sec:instability}. Some examples of the short-term evolution following a localised crustal heating are illustrated in section \ref{sec:heating}, with a view to applications to magnetar outbursts (section \ref{sec:deepheating}) and to surface heating in pulsars (section \ref{sec:surfheating}). Discussion and conclusions follow in section \ref{sec:conclusions}.

\section{The model} \label{sec:equations}
\subsection{Input physics and evolution equations}
The NS crust comprises a Coulomb lattice in which nuclei have negligible motion. Hence, the crustal currents are produced entirely by the flow of electrons, which form a highly relativistic and strongly degenerate Fermi gas. Still, their mean velocity is typically only a tiny fraction of the speed of light. We can therefore resort to the (non-relativistic) electron magneto-hydrodynamics (eMHD) approximation in treating the crustal dynamics. The evolution of the magnetic field $\mathbf{B}$ in the crust is described by the induction equation that, taking into account also the effects of thermal coupling, can be written in the form
\begin{equation}
 \frac{\partial\mathbf{B}}{\partial t} = -c\,\nablab\times\left[\boldsymbol{\sigma}^{-1}\cdot\mathbf{J} + \boldsymbol{G}\cdot\nablab T-\nablab\mu/\ee\right] \label{eq:Induction}
\end{equation}
where the term in square brackets is the electric field $\boldsymbol{E}$ as given by generalised Ohm's law. Here, $c$ is the speed of light, $\ee$ the electron charge, $\boldsymbol{\sigma}$ and $\boldsymbol{G}$ are the electric conductivity and the thermopower tensors, and $\mu$ is the electron chemical potential. The latter, for a degenerate relativistic Fermi gas, depends only on the density, $\mu= c\hslash (3\pi^2n)^{1/3}$ where $\hslash$ is the reduced Planck constant. Neglecting the displacement current, the electron current is given by $\mathbf{J} = c\,\nablab\times\mathbf{B}/4\pi$. 

Assuming the temperature of the crust to be well below the electron degeneracy temperature, but above the ion plasma temperature,
scattering of electrons can be described in terms of an energy-dependent relaxation time $\tau$ \cite[e.g.][]{ziman1972principles, 1980SvA....24..425U},
and the electron conductivity can then be approximated as
\begin{equation}
 (\boldsymbol{\sigma}^{-1})_{ij} = \sigma^{-1}\delta_{ij} + \frac{\epsilon_{ijk}B_k}{\ee cn}
\label{eq:sigma}\end{equation}
where the symmetric part is
\begin{equation}
 \sigma = \ee^2c^2\,\frac{n\tau(\mu)}{\mu}
\end{equation}
and the anti-symmetric part represents the Hall effect.

The thermopower can be calculated using the Mott formula, and in general has an isotropic part and a part proportional to the conductivity tensor. In our model we include only the isotropic part (the so-called \emph{Seebeck term}), which is responsible for the Biermann battery effect,
\begin{equation}
 \boldsymbol{G} = \ee\left.\frac{\partial\boldsymbol{\sigma}^{-1}}{\partial\mu}\right|_T\cdot\boldsymbol{k}\simeq -\frac{\pi^2k_B^2T}{\ee\mu}\delta_{ij}
\end{equation}
where the approximate equality is obtained further assuming electrons to form perfect Fermi gas. Here, $\boldsymbol{k}$ is the thermal conductivity tensor, that is taken to be proportional to $\boldsymbol{\sigma}$, according to the Wiedemann-Franz law
\begin{equation}
 \boldsymbol{k} = \frac{\pi^2k_{\rm B}^2T}{3\ee^2}\,\boldsymbol{\sigma}\label{eq:wf}
\end{equation}
where $k_{\rm B}$ is Boltzmann's constant.

The evolution of temperature is, in turn, governed by the heat equation 
\begin{equation}
 C_V\frac{\partial T}{\partial t} = - \nablab\cdot\left( T\boldsymbol{G}\cdot\mathbf{J} - \boldsymbol{k}\cdot\nablab T - \frac{\mu}{\ee}\mathbf{J}\right) + \mathbf{E}\cdot\mathbf{J}+ N_\nu \label{eq:heat}
\end{equation}
where $C_V$ is the heat capacity (per unit volume) of the crust, $N_\nu$ is the neutrino emissivity due to weak processes, and the term in round brackets is the electron energy flux.

Although general-relativistic effects can be accounted for with no inherent difficulty in equations (\ref{eq:Induction}) and (\ref{eq:heat}), see e.g. \citet{2009A&A...496..207P} and \citet{2013MNRAS.434..123V}, they are not included here. The reason for this is twofold. First, given the small thickness of the crust, they are of limited importance and will not change our results qualitatively and, second, a proper general-relativistic treatment impacts on the boundary conditions which are imposed on the evolution equations (see section \ref{bcondit}). While this poses no serious problem in 2D, it becomes quite troublesome in 3D. Equations (\ref{eq:Induction}) is analogous to the one solved in \citet{2014MNRAS.438.1618G} (with the addition of thermocoupling terms) and \cite{2013MNRAS.434..123V}, where (a different version of) equation \ref{eq:heat} was included as well.

In order to simplify the equations, all physical quantities are henceforth expressed in terms of values typical of the outer crust in a magnetar. In particular, the temperature, magnetic field, relaxation time and chemical potential are normalised to
$T_0 = \SI{e8}{\kelvin}$,
$B_0 = \SI{e14}{\gauss}$,
$\mu_0=\SI{2.9e-5}{\erg}$,
and $\tau_0=\SI{9.9e-19}{\second}$.
This implies that the reference values for the number density $n$ and the conductivity $\eta = c^2/(4\pi\sigma)$ are $n_0 \simeq 2.6\times10^{34}$cm$^{-3}$
and $\eta_0 \simeq 3.9\times10^{-4}$cm$^2$s$^{-1}$. It is also useful 
to introduce four length scales, which are relevant to the electron dynamics,
\begin{align}
  \lambda &= \left(\frac{k_{\rm B}T_0}{4\pi n_0\ee^2}\right)^{1/2} & \text{Debye length} \\
  d &= \left(\frac{\mu_0}{4\pi n_0\ee^2}\right)^{1/2} & \text{skin-depth} \\
  L &= \frac{\mu_0}{\ee B_0} &\text{Larmor radius} \\
  l &= c\tau_0 & \text{mean free path,}
\end{align}
supplemented by the star radius $R_\star=\SI{10}{\kilo\meter}$ as the macroscopic  length scale. Furthermore, the ohmic time $\tau_{O}=R_\star^2/\eta_0\simeq\SI{8e7}{\year}$ is taken as the reference time scale, and $C_{V0} = k_{\rm B}n_0$ as the scale for the heat capacity.

The evolution equations to be solved for $\mathbf{B}$ and $T$ become then
\begin{align}
  \dfrac{\partial\mathbf{B}}{\partial t}
  =\,\Se\,\nablab&\left(\frac{1}{\mu}\right)\times\nablab T^2
    + \Ha\,\nablab\times\left[\frac{1}{\mu^3}\mathbf{B}\!\times\!(\nablab\times\mathbf{B})\right]+ \notag 
    \\&- \nablab\times\left[\frac{1}{\tau\mu^2}\nablab\times\mathbf{B}\right]\label{eq:Hall}\\
   \notag\frac{1}{\Ro}\frac{C_V}{T}\dfrac{\partial T^2}{\partial t} =& \nablab\cdot\left(\tau\mu^2\boldsymbol{\chi}\cdot\nablab T^2\right)
  + \frac{\Pe}{\Se}\frac{|\nablab\times\mathbf{B}|^2}{\tau\mu^2}+
 \\&+ \Pe\,\mu(\nablab\times\mathbf{B})\cdot\nablab\left(\frac{T^2}{\mu^2}\right) +\frac{1}{\Ro} N_\nu\label{eq:T}\end{align}
where we defined
\begin{equation}\chi_{ij} = \frac{\delta_{ij} 
    + \Ha^2(\tau/\mu)^2B_iB_j
    - \Ha(\tau/\mu)\epsilon_{ijk}B_k}
    {1 + \Ha^2(\tau/\mu)^2|\mathbf{B}|^2}
\end{equation}
and introduced the adimensional numbers
\begin{align}
    \Ha &= l/L \simeq 50 &\textit{Hall}\\
    \Se &= \frac{\pi^2Ll\lambda^4}{2d^6} \simeq 0.05 &\textit{Seebeck}\\
    \Pe &= \frac{3d^2}{Ll} \simeq 6\times10^{-5} &\textit{Peclet}\\
    \frac{1}{\Ro} &= \sqrt{\frac{3}{2\pi^2}\frac{\Pe}{\Se\,\Ha^2}}  %\frac{3d^4}{\pi^2l^2\lambda^2} 
    \simeq 3\times10^{-4}&\textit{Roberts}.
\end{align}
Performing adimensionalisation with these scales, the neutrino emissivity is expressed in units of $N_\nu^0=8\pi\ee^2 c^2 R^2 \tau_0/\mu_0k_bT_0=\SI{1.3e14}{\erg\per\second\per\cubic\centi\meter}$.

The large thermal conductivity of the crust, which is reflected in the fact that $\Ro \gg 1$, means that on the timescale of magnetic evolution the term dependent on the heat capacity of the crust is subdominant. Rather than using a detailed microphysical model for $C_V$, we therefore simply take $C_V = \tau\mu^2T$, which implies a constant effective thermal diffusivity throughout the crust.
Under this assumption, equation (\ref{eq:T}) depends on temperature only through $T^2$, which proves to be an advantageous feature for numerical implementation.

Under the eMHD approximation, electrons and protons in the crust have equal, time-independent number density $n$. We will assume that the crust is spherically symmetric, so that $n$ is a function of the radius $r$ alone. With our definition of the chemical potential (see above), this implies that $\mu$ depends on $r$ and we take 
\begin{equation}
  \mu(r) = \left(1 + \frac{1 - r}{0.0463}\right)^{4/3},\label{n_profile}
\end{equation}
following \citet{2014MNRAS.438.1618G}; $\mu(r)$ increases from unity at the outer boundary ($r=1$) to $\simeq4.6$ at the inner boundary ($r=0.9$).
Moreover, we also assume that $\tau$ is a function of $r$ only and, in particular, we take $\tau\equiv1$.

\added{The density profile corresponds to the crust model with impurity parameter $Q\simeq3$ of \citet{2004ApJ...609..999C}. The assumption of taking the relaxation time $\tau$ to be independent on temperature is adequate in the lower crust, while it is just an approximation in the upper crust, where scattering is dominated by phonons \citep{transport}. We, nevertheless, note that taking $\tau\equiv1$, the conductivity in the upper crust corresponds to the phonon conductivity at a realistic temperature, $T\approx\SI{e8}{\kelvin}$. We assume an Fe-Ni crust, without accounting for chemical composition stratification.}

The emission of neutrinos in the crust is due to a large variety of reactions. In this work, the four dominant contributions are taken into account, namely phonon decay, neutrino pair production, neutrino bremsstrahlung and neutrino synchrotron emission
\begin{align}
    N_\nu(n,T,\mathbf{B})= &N_{\text{ph}}(n,T)+N_{\text{pair}}(n,T)+\\
    \nonumber
     & N_{\text{bre}}(n,T)+N_{\text{syn}}(n,T,\mathbf{B})\,.
\end{align}
We make reference to \citet{bremsstrahlung} for neutrino pair bremsstrahlung decay and to \citet{phonon} for phonon decay. A complete review can be found in \citet{2001PhR...354....1Y}. These papers provide fitting formulae for numerical evaluation, that were implemented in our code.

\subsection{Boundary conditions}
\label{bcondit}

Solution of the evolution equations (\ref{eq:Induction}) and (\ref{eq:heat}) requires boundary conditions which reflect a number of physical prescriptions at the core-crust interface and at the star surface. We assume that all magnetic flux has been expelled from the  superconducting core. This requires that the normal component of the magnetic field and \replaced{the tangential components of the current}{the tangential component of the electric field} must vanish at the core-crust boundary, $r=r_c$. \added{The latter results in a nonlinear boundary condition for the magnetic field due to the presence of the Hall term. Nevertheless, this contribution is negligible near the bottom of the crust due to the high electron density \citep[see][]{2004MNRAS.347.1273H}, and can hence be neglected. This allows to write the boundary conditions in terms of the radial magnetic field and of the  tangential component of the current $J_t$,}
\begin{equation}
    B_r(r_c)=0\qquad J_t(r_c)=0.\label{eq:typeI}
\end{equation}

We assume that the electrical conductivity in the magnetosphere is negligible in comparison with that of the crust, and therefore match the field at the crust outer boundary to  a potential one. \added{This can be achieved in a very natural way by exploiting the spectral nature of our code, which decomposes the field using the spherical harmonics $Y_\ell^m(\theta,\phi)$ as the basis (see section \ref{sec:numerics}), after introducing a poloidal-toroidal decomposition
\begin{equation}\mathbf{B}=\nablab\times\nablab\times(\mathbf{r}B_\text{pol})+\nablab \times(\mathbf{r}B_\text{tor}).\label{eq:pol-tor}\end{equation}
In such a representation each mode of a potential field is purely poloidal and such that {$(B_\text{pol})_\ell^m\propto r^{-(\ell+1)}$}. Hence, the boundary condition can be met by requiring that 
\begin{equation}\begin{aligned}
&\left.\frac{\partial}{\partial r}(B_\text{pol})_\ell^m+\frac{\ell+1}{r}(B_\text{pol})_\ell^m\right|_{r=R_\star}\!\!\!\!=0;\\
&\left.(B_\text{tor})_\ell^m\right|_{r=R_\star}\!\!=0.
\end{aligned}\label{BC_Bext}\end{equation}}
%\begin{equation}
%\nablab\times\mathbf{B}\vert_{R_\star}=0\,.\label{BC_Bext}
%\end{equation} 

The core conducts heat even more efficiently than the crust, and is therefore approximately isothermal.
It cools by neutrino emission according to the equation
\begin{equation}
\label{eq:core-cool}
\frac{\partial T_{c}}{\partial t}=-\frac{N_c(T)}{C_c}
\end{equation}
where $C_c$ is the core specific heat and $N_c(T)=N_0T^k$ the neutrino emissivity of the core. The star long-term thermal evolution is governed by eq. (\ref{eq:core-cool}) once the neutrino emissivity is specified. Our model uses a standard slow-cooling scenario with $k=8$, $C_c=\SI{e20}{\erg\per\second\per\kelvin^2}$, $N_0=\SI{e21}{\erg\per\cubic\centi\meter\per\kelvin^8}$ \citep{2004ApJS..155..623P}. 

The surface temperature is controlled by the properties of the thermal blanketing envelope. This layer is geometrically very thin, but hosts a large temperature gradient. Thus, the widespread approach is to treat it separately, using a plane-parallel approximation to obtain a relation between the temperature at the bottom of the envelope $T_b$ (that is, the temperature of the top of the crust) and the surface temperature $T_s$ \citep{1983ApJ...272..286G}. %, to relate with the surface temperature gradient
Assuming that no energy gains or losses occur in the envelope, the temperature gradient at the top of the crust is given by \citep{tsuruta}
\begin{equation}
 - (\boldsymbol{k}\cdot\nablab T)\cdot{\hat{\mathbf{r}}} = \sigma_{\text{\sc sb}} \,T_s^4(T_b,\mathbf{B})\,,
\end{equation}
where the left-hand side is evaluated at the top of the crust
and $\hat{\mathbf{r}}$ is the radial unit vector.
We have chosen the form
\begin{equation}\label{eq:tsurf}
T_s(T_b, g, \mathbf{B})=T_s^{(0)}(T_b, g)\,\mathcal{X}(T_b, \mathbf{B})
\end{equation}
where $g$ is the gravitational acceleration at the surface. We used the expressions for $T_s^{(0)}$ as calculated in \citet{1983ApJ...272..286G} for an iron envelope neglecting magnetic fields, and the magnetic correction $\mathcal{X}(T_b,\mathbf{B})$ obtained in \citet{envelopes}.

\subsection{Numerical implementation}
\label{sec:numerics}

Equations (\ref{eq:Hall}) and (\ref{eq:T}) were solved in three dimensions using a suitably modified version of the code {\sc parody}, which was originally developed by \citet{1998E&PSL.160...15D} and \citet{2008GeoJI.172..945A}. A version of the same code, which did not include the thermo-magnetic coupling, was first used to investigate the magnetic field evolution in NSs in \citet{2015PhRvL.114s1101W}. 
The code is pseudo-spectral: it uses a finite grid in the radial direction and an expansion in spherical harmonics $Y_\ell^m(\theta,\phi)$ for the angular part. The NS crust is assumed to be a perfect spherical shell. The time-stepping algorithm is Crank-Nicholson for the Ohmic diffusion, backward-Euler for the isotropic part of the thermal diffusion, and Adams-Bashforth for all other terms.

We typically use 128 radial grid points, and spherical harmonics up to degree $\ell\approx 100$, obtaining a typical resolution of $\lesssim\SI{100}{\meter}$ on the surface. Parallelisation is implemented using {\sc{mpi}} and the code is run on a cluster of CPUs. Work is distributed in such a way that each thread takes care of a spherical shell containing $N_r\mod{N_\text{cores}}$ points, where $N_r$ is the radial grid size. In order to compute space derivatives within our finite difference scheme in each thread, a single shell should contain at least four grid points, hence to achieve the desired resolution the code is typically run on $\lesssim 32$ cores.

%%%%%%%%%%%%%%%%%%%%%%%%%%%%%%%%%%%%%%%%%%%%%%%%%%%%%%%%%%%%%%%%%%%%%%%%%%%%%%%%%%%%%%%%%%%%%%%%%%%%%%%%%%%%%
\section{Study cases}\label{sec:cases}

In order to validate the code and provide comparisons with previous works, we first address the problem of the secular magneto-thermal evolution of highly magnetised, isolated NSs. The magnetic evolution follows two different timescales, the Hall and Ohm ones \citep{1992ApJ...395..250G},
\begin{align}\tau_H&=\dfrac{4\pi n_0 e R^2}{cB_0}\approx \SI{e4}{\year}\\ \tau_O&=R^2/\eta_0\approx \SI{e7}{\year}.
\end{align}
Magnetic field reconfiguration occurs on the Hall timescale, when small scale structures are formed by the action of the Hall term, while on the Ohm one, dissipation takes place. Long-term thermal evolution also occurs
on a time $\lesssim\tau_O$ \cite[see e.g.][for a review]{transport}. A 3D approach is particularly suited, and indeed necessary, to follow the formation and evolution of small-scale structures in the Hall phase. 

\subsection{Neutron star magneto-thermal evolution}

In order to set the initial\footnote{We remark that throughout the work the initial time is set in correspondence to the superfluid transition, which typically occurs a few years later than the formation of the proto NS.} magnetic configuration for our simulations, we followed the widespread approach of confining the field in \added{simple,} large-scale structures \citep{rudiger2013magnetic}. In particular, we selected a \added{force-free $B$-field matching our boundary conditions, with both non-zero poloidal ($\ell=1$, $m=0$) and toroidal ($\ell=2$, $m=0$) components. For such a field, the components of equation (\ref{eq:pol-tor}) take the form $B_\ell^m\propto Y_\ell^m(\theta,\phi)\,\zeta_\ell(r)/r$, where $\zeta_\ell$ is a linear combination of spherical Bessel functions of degree $\pm\ell$, constructed in such a way to obey the boundary conditions \citep[see][for a full derivation]{1957ApJ...126..457C}}. We stress that the evolution of poloidal/toroidal components is strongly coupled by the action of the Hall term, that can transfer energy both ways between them \citep{2007A&A...470..303P}.
The initial temperature profile is assumed to be a constant, $T(r, t=0)\equiv\SI{e8}{\kelvin}$, but we note that the overall evolution is virtually independent on this choice. This is due to the fact that the term $\propto{\partial T}/{\partial t}$ in equation (\ref{eq:T}) is suppressed by a factor $\Ro^{-1}\approx 10^{-4}$, so that the temperature rapidly achieves a quasi-steady state.

\begin{figure}
    \centering
    \includegraphics[width=.46\textwidth]{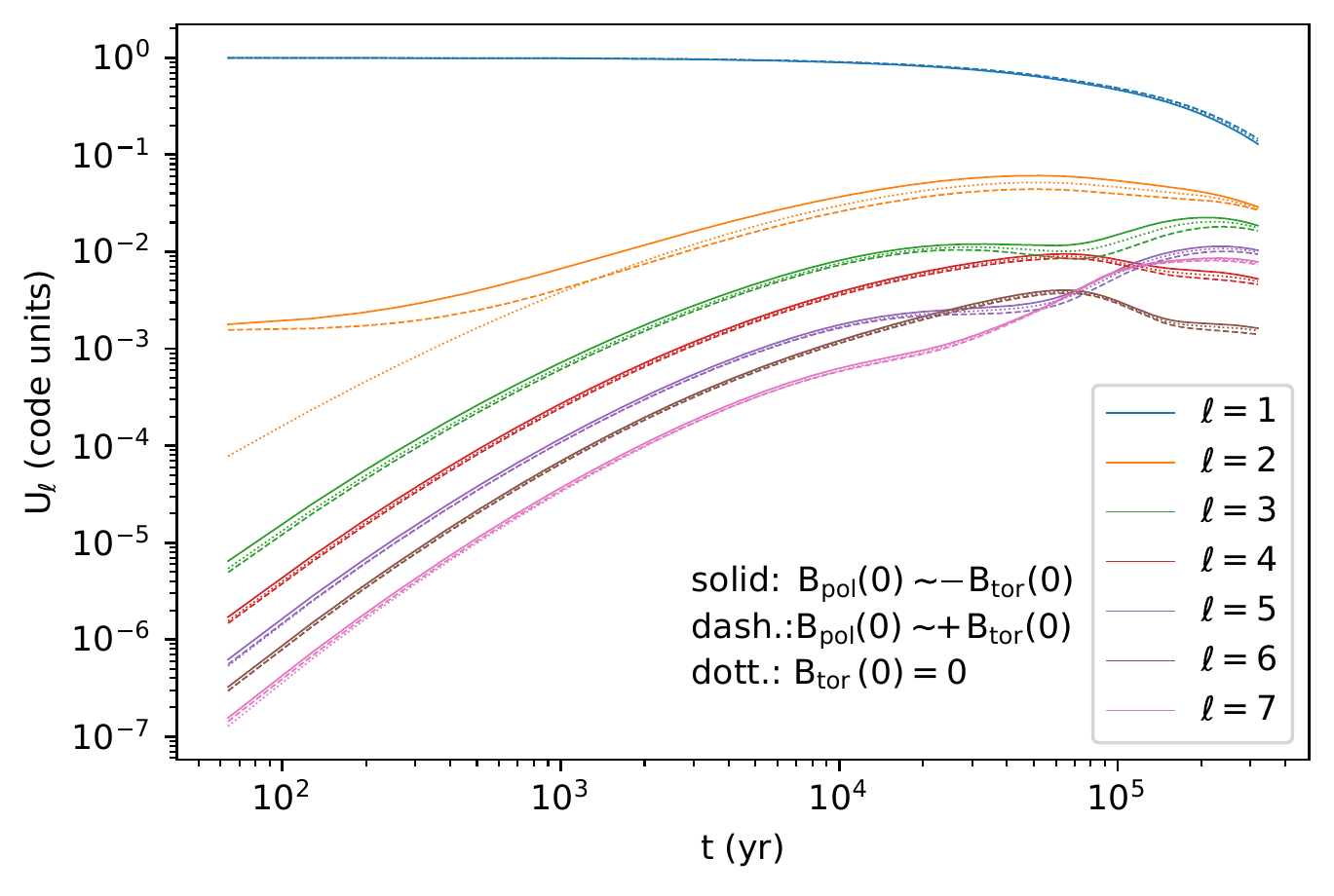}
    \caption{Time evolution of the energy in the first seven $\ell$ modes for three typical initial configurations of the field: a purely poloidal field (solid) and two cases with an added toroidal field of opposite polarity (dashed and dotted).}
    \label{fig:evo_spec}
\end{figure}

The evolution of the B-field over a few Hall timescales is shown in Figure \ref{fig:evo_spec} for three different initial magnetic configurations: a purely dipolar field ($B_\text{pol}(0)\approx\SI{e14}{\gauss}$, $B_\text{tor}(0)=0$) and a field with poloidal and toroidal components of the same order but opposite relative polarity ($B_\text{pol}(0)\approx\pm B_\text{tor}(0)\approx\SI{e14}{\gauss}$). Our simulations confirm the previous finding that the magnetic field evolves towards the so-called \emph{Hall attractor} \citep{2014PhRvL.112q1101G}, where the magnetic field tends to reach a configuration dominated by the modes $\ell=1, 2, 3, 5, 7$ (see again Figure \ref{fig:evo_spec}). The dominance of odd modes with respect to the nearby even ones is a general feature of the Hall attractor. We remark that for a better comparison with previous works \citep{2013MNRAS.434..123V, 2011ApJ...740..105T}, we chose initial conditions that are essentially axisymmetric. Our results show that initially axisymmetric configurations tend to maintain their symmetry as they evolve.

\begin{figure*}
    \centering
%\subfigure{\includegraphics[width=.3\textwidth, trim= 90 35 40 32,clip]{Br_mer1.png}}~~
%\subfigure{\includegraphics[width=.3\textwidth, trim= 90 35 40 32,clip]{Bt_mer1.png}}~~
%\subfigure{\includegraphics[width=.3\textwidth, trim= 90 35 40 32,clip]{Bp_mer1.png}}\\
%\subfigure{\includegraphics[width=.3\textwidth, trim= 90 35 40 32,clip]{Br_mer2.png}}~~
%\subfigure{\includegraphics[width=.3\textwidth, trim= 90 35 35 32,clip]{Bt_mer2.png}}~~
%\subfigure{\includegraphics[width=.3\textwidth, trim= 95 35 40 32,clip]{Bp_mer2.png}}\\
\includegraphics[width=.9\textwidth]{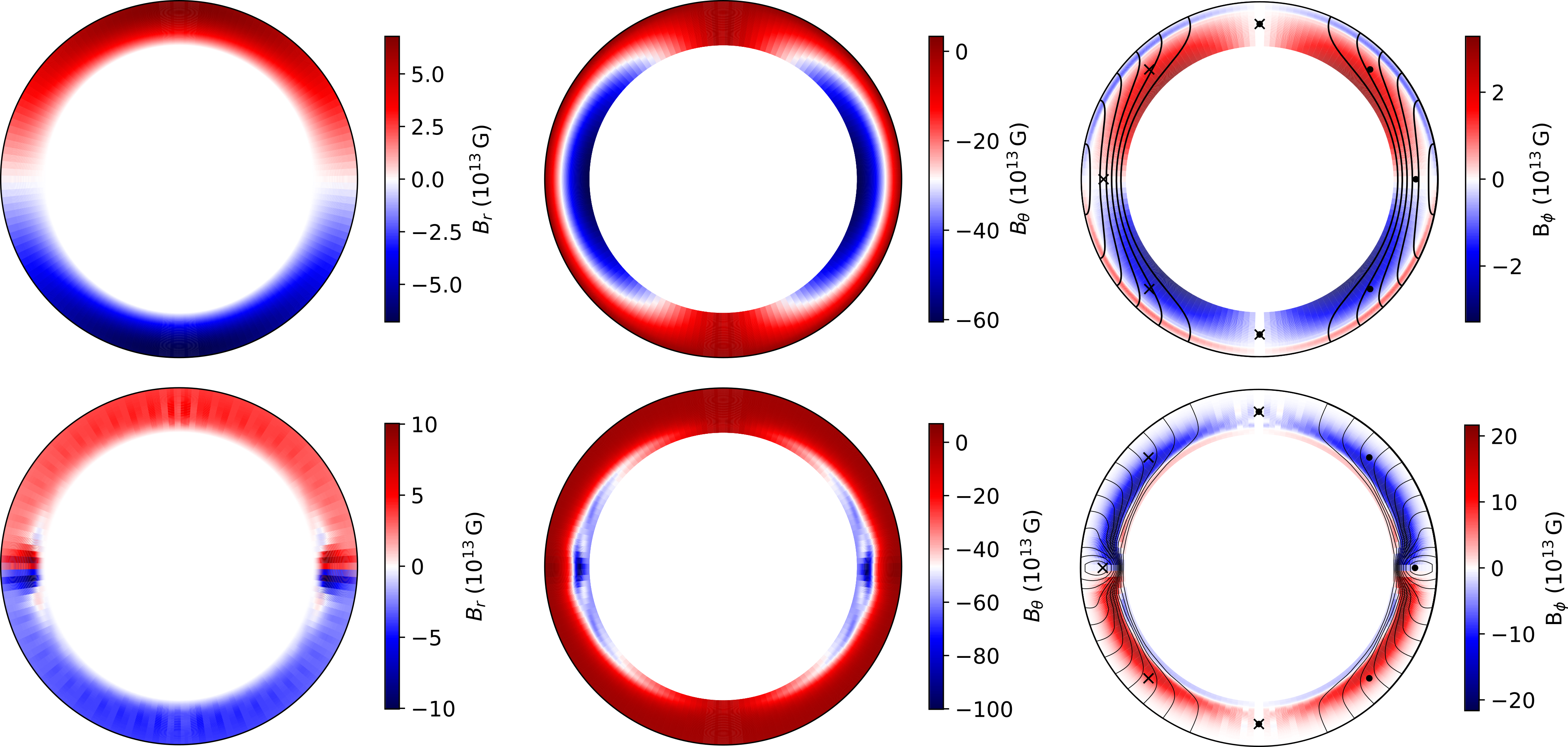}
\caption{A meridional cut of the crust along the prime meridian ($\phi=0$) showing the the magnetic field components $B_r$, $B_\theta$ and $B_\phi$ (from left to right) at the start (top row) and after $\SI{3e4}{\year}\approx\tau_H$ (bottom row) for the run with $B_\text{pol}(0)\sim+ B_\text{tor}(0)$. \added{The plots for the $\phi$ component also show the field lines of the poloidal field.} Here and in all figures where relevant, the crust thickness is enhanced by a factor $4$ for better visualisation.}
\label{fig:Bfled_mer}
\end{figure*}
The components of the magnetic field in spherical coordinates, $B_r$, $B_\theta$ and $B_\phi$, at the beginning of the simulation ($t=0$) and at $t=\SI{3e4}{\year}\simeq \tau_H$  are shown in figure \ref{fig:Bfled_mer} for the  case with $B_\text{pol}(0)\approx+ B_\text{tor}(0)$. A general feature of the magnetic field is the appearance of an equatorial structure in which the field is stronger \citep{2016PNAS}, and of small scale structures due to the Hall term.

As already mentioned, the temperature distribution tends to follow the magnetic field. The structure of the Hall attractor, in which an equatorial current ring forms, is reflected in a hotter equatorial region. Moreover, formula (\ref{eq:wf}) implies that heat tends to be transported preferentially along the field lines.
Hence, the equatorial region is hotter not only because of higher dissipation, but also because heat is trapped by the closed field lines appearing in that region. Figure \ref{fig:temperature} shows a typical case, that is representative---at least qualitatively---of all our nearly-axisymmetric runs.
\added{We note that, owing to the dependence of the properties of the heat blanketing envelope on the geometry of the magnetic field, the observable surface map can be quite different from the one on the top of the crust. As an example, the last panel of Figure \ref{fig:temperature} shows the surface temperature for the very same case: the overall topology is quite different, as the equatorial belt is not just hotter, but instead shows a colder ring at the very equator \citep[see e.g. the recent results in][in which a similar behaviour is discussed in a 3D stationary framework]{2019hepr.confE..59K}.
\begin{figure}
    \centering
    \includegraphics[width=.48\textwidth]{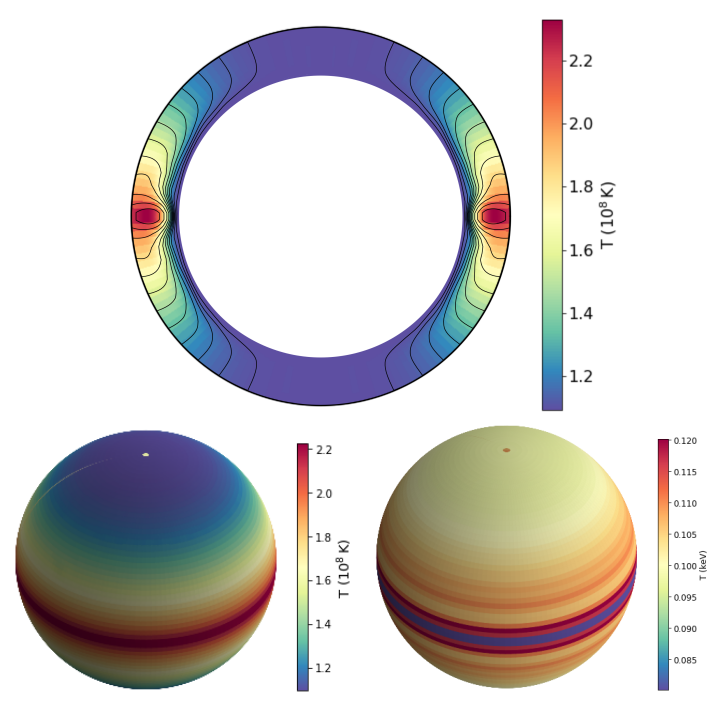}
    \caption{Temperature maps at $t=\SI{3e4}{\year}\approx\tau_H$ for a case with $B_0^\text{pol}\approx+B_0^\text{tor}\approx\SI{e14}{\gauss}$. {\emph{Top}}: meridional cut,  with the field lines of the poloidal component superimposed; \emph{Bottom left}: temperature at the top of the crust (i.e. under the heat blanketing envelope); \emph{Bottom right}: surface temperature according to equation (\ref{eq:tsurf}), showing how the envelope can change the very topology of the temperature distribution.}
    \label{fig:temperature}
\end{figure}
}
Even though the various features are on a large scale, they exhibit a smaller scale --yet well resolved-- structure, due to the Hall term.

%\begin{figure}
%    \centering
%    \subfigure{\includegraphics[width=.3\textwidth]{T_mer_Hall_lines.png}}\\
%    \subfigure{\includegraphics[width=.3\textwidth]{T_surf_Hall.png}} 
%    \caption{Temperature maps at $t=\SI{3e4}{\year}\approx\tau_H$ for a case with %$B_0^\text{pol}\approx+B_0^\text{tor}\approx\SI{e14}{\gauss}$. {\emph{Top}}: meridional cut, shown with the lines of the poloidal field; \emph{Bottom}: temperature at the top of the crust (i.e. under the heat blanketing envelope).}
%    \label{fig:Tmer}
%\end{figure}

\subsection{Magnetars and eMHD instabilities}\label{sec:instability}

 As already noted in \citet{2020arXiv200103335G}, the presence of a strong toroidal field can trigger a \emph{resistive tearing} eMHD instability \citep{2014PhPl...21e2110W}. This instability, even when starting from an initial condition that is essentially symmetric, produces non-axisymmetric small-scale magnetic structures, that, due to Joule dissipation, translate into localised heat deposition. A strong toroidal component in the star crust passes undetected and is invoked to explain the observed activity in the so-called \emph{low-B} magnetars, i.e. sources with a dipole field comparable to that of the radio-pulsar population (see section \ref{sec:discussion} for further details).

\begin{figure}
    \centering
    \includegraphics[width=.46\textwidth]{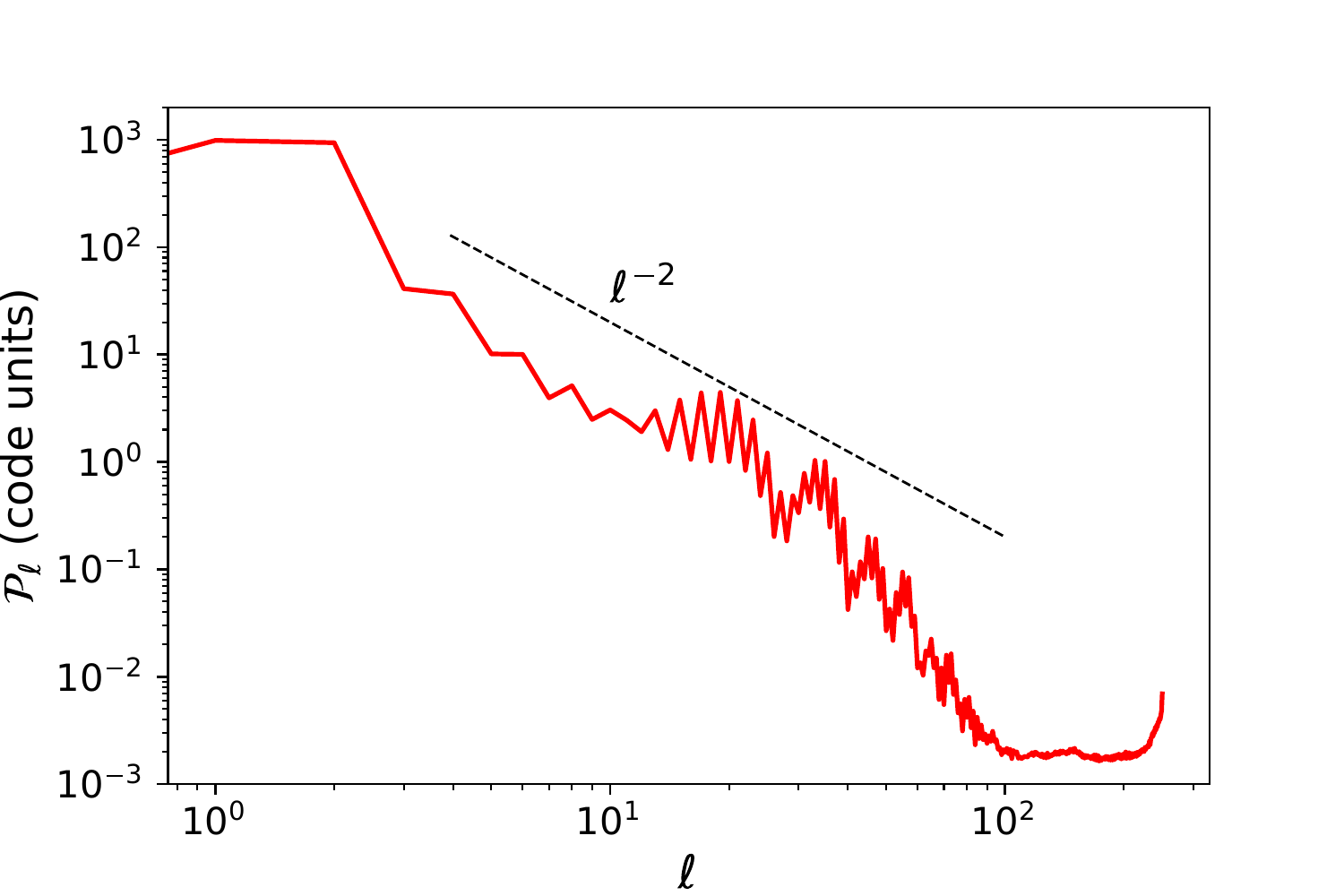}
    \caption{Power spectrum (in code units) of the $\ell$ modes at $t\simeq\SI{e4}{\year}$. On top of the wavy profile privileging odd modes, typical of the Hall attractor, a more complex pattern at short wavelengths reflecting the instability is visible. The slope $\ell^{-2}$, obtained from scaling relations for Hall turbulence in \citet{1992ApJ...395..250G}, is shown for reference.}
    \label{fig:spectra}
\end{figure}

To explore better this issue, we ran a simulation assuming an $\ell=1$, $m=0$ initial magnetic field with a poloidal field $B_\text{pol}(0)\approx\SI{e14}{\gauss}$ and a toroidal one $B_\text{tor}\approx\SI{4e15}{\gauss}$. Given the nature of the solution we are looking for, the resolution for this case was improved to $\ell_\text{max}=250$, corresponding to cells of a few tenths of meters on the surface. Indeed, an instability is triggered after about a Hall time $\tau_H$. The spectrum of all the $\ell$ modes at $t\simeq\SI{e4}{\year}$ (Figure \ref{fig:spectra}) exhibits the characteristic features of the Hall attractor: even modes are suppressed with respect to the nearby odd ones up to $\ell\lesssim100$ and this produces the typical wavy profile. However, the onset of an instability is marked by the appearance of well-resolved structures that form up to $\ell\simeq100$, with a complex structure of secondary peaks on top of the Hall structure. The flatness of the spectrum at high $\ell$ guarantees that the instability is of physical and not numerical origin. The slight increase at very high $\ell$ is due to numerical aliasing. The spectrum of $m$ modes, on the other hand, is sharply peaked towards zero and hence is not shown. 

\begin{figure}
    \centering
    \includegraphics[width=.46\textwidth]{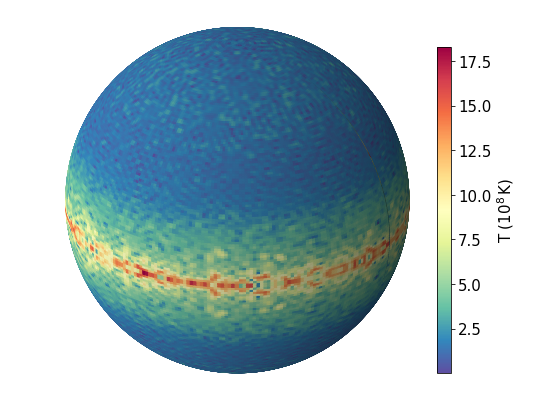}
    \caption{Temperature at the top of the crust showing the formation of a hotter equatorial belt with a small-scale, yet numerically resolved, pattern that reflects the eMHD instability. Note that this is not the surface temperature.}
    \label{fig:instaT}
\end{figure}

In the small structures, the magnetic field can reach values up to $\sim\SI{2e16}{\gauss}$ and this drives a local temperature increase, as shown in figure \ref{fig:instaT}. Such strong fields generate high magnetic stresses in the crust. As a gauge to determine whether such stresses are strong enough to lead to a crustal failure, we compared them to the maximum mechanical yield of the crust through the von Mises criterion \citep[see e.g.][]{2012ApJ...750L...6P,2019MNRAS.486.4130L},
\begin{equation}
\sqrt{\bar M_{ij}\bar M^{ij}}%=\sqrt{M_{r\theta}+M_{r\phi}+M_{\theta\phi}\phantom{\bar I}\!\!}\,
\gtrsim \tau_\text{max}(n,T)\label{eq:vM}
\end{equation}
where $\bar M_{ij}$ is the traceless part of the magnetic stress tensor $M_{ij}=B_iB_j/4\pi$. 
\citet{2010MNRAS.407L..54C} derived estimates for $\tau_\text{max}$ by means of molecular dynamic simulations and elucidated the strong dependence of the breaking stress on temperature. \added{In our calculation we used the fit they  provide for the maximum crustal yield, \begin{equation}
    \tau_{max}=\left(0.0195-\frac{1.27}{\Gamma-71}\right)n_i\frac{Z^2\ee^2}{a}
\end{equation} 
where $\Gamma=Z^2\ee^2/ak_BT$ is the classical Coulomb coupling parameter, $n_i$ is the ion density, $a=(4\pi n_i/3)^{-1/3}$ is the ion sphere radius and, following  \citet{2007PhRvE..75f6101H}, we took $Z=29.4$ as the mean ion charge in the Fe-Ni crust.} Since $\tau_\text{max}$ decreases at higher $T$, a 3D, coupled magneto-thermal code provides the most accurate way to investigate the onset of crustal failures in magnetars.

\begin{figure}
    \centering
    \includegraphics[width=.48\textwidth]{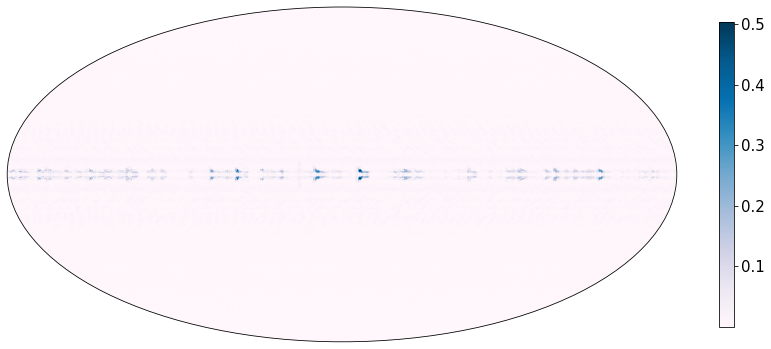}
    \caption{Ratio between magnetic stresses and maximum yield during the instability, shown at the radius where is attains its maximum value, close to half of the crust depth.}
    \label{fig:stress}
\end{figure}

Figure \ref{fig:stress} shows the ratio between the magnetic and the breaking stress in our simulation after a time $\SI{6.2e3}{\year}\lesssim\tau_H$. The map refers to the region of the crust where the ratio is maximum, at about half the crust depth. The magnetic stress reaches values up to $\sim 50\%$ of the maximum yield in our simulation so that von Mises criterion for crustal yielding is likely to be fulfilled. Crustal failures can therefore be triggered in our simulation because of the large magnetic stress coupled with the heating produced by magnetic dissipation which significantly rises the temperature in the equatorial region, thus lowering the breaking stress. The instability lasts for some $\SI{1000}{\year}$ before it is damped by dissipation.
This directly concerns magnetar activity, since magnetically induced crustal failures (``starquakes'') are thought to be responsible for magnetar bursts and outbursts \citep{2012ApJ...750L...6P}.

\added{We conclude this section noticing that the use of the von Mises criterion as expressed by equation (\ref{eq:vM}), albeit widespread in the literature, should be taken with some care. In fact, it does not take into account the effects of the enormous gravity, that tends to inhibit any radial displacement \citep[see e.g.][]{2008CQGra..25k4049H}. However, since the resistive tearing instability arises as a consequence of the presence of a strong toroidal field, our result is not much affected even setting to zero all radial shear terms (the maximum stress-to-yield ratio decreases from $\sim 50$\% to $45$\%). Nevertheless, only a consistent, non-local calculation which takes into account the global hydrostatic structure of the crust, could unambiguously solve the issue.}

{}

\section{Localised heating in NS crust}\label{sec:heating}
In order to fully exploit the three-dimensionality of our code, we investigated models in which a localised heat source is present in the NS crust. This is accounted for by adding a term $\dot H/\mathcal{V}_\text{heated}$ to equation (\ref{eq:heat}), describing the heat injection rate per unit volume. In particular, we consider two cases: (i) localised heat deposition in the deep crustal layers, and (ii) heating of the star's external layers. Although no direct application to real astrophysical sources will be attempted, these two models are of interest in connection with the evolution of magnetar outbursts and the X-ray emission from radio-pulsars.

\subsection{Heating in the deep crust}\label{sec:deepheating}

As already mentioned, magnetar activity is believed to be associated with crustal failures (see Section \ref{sec:instability}). However, the crustal dynamics in such events is little explored as yet, owing to its inherent complexity \cite[e.g. the crust may flow plastically,][]{2016ApJ...824L..21L}. Such a study is beyond the capability of our code, which does not incorporate a description of the motion of crustal matter. 

As a minimal model to address the physics of crustal failures, and in particular the way in which heat is transported to the surface, we therefore performed a simulation in which energy is injected during a short time interval in  a localised region of the crust, much in the same way as in \citet{2012ApJ...750L...6P} but exploiting our fully 3D approach.
As the background state, we take a NS with an initial field $B_\text{pol}\approx\SI{e12}{\gauss}$ and $B_\text{tor}\approx\SI{e13}{\gauss}$ that has been consistently evolved for a Hall time. This high toroidal field configuration was chosen in the spirit of the results of section \ref{sec:instability} and mimics a low-B magnetar.

\begin{figure}
    \centering
    \includegraphics[width=.45\textwidth]{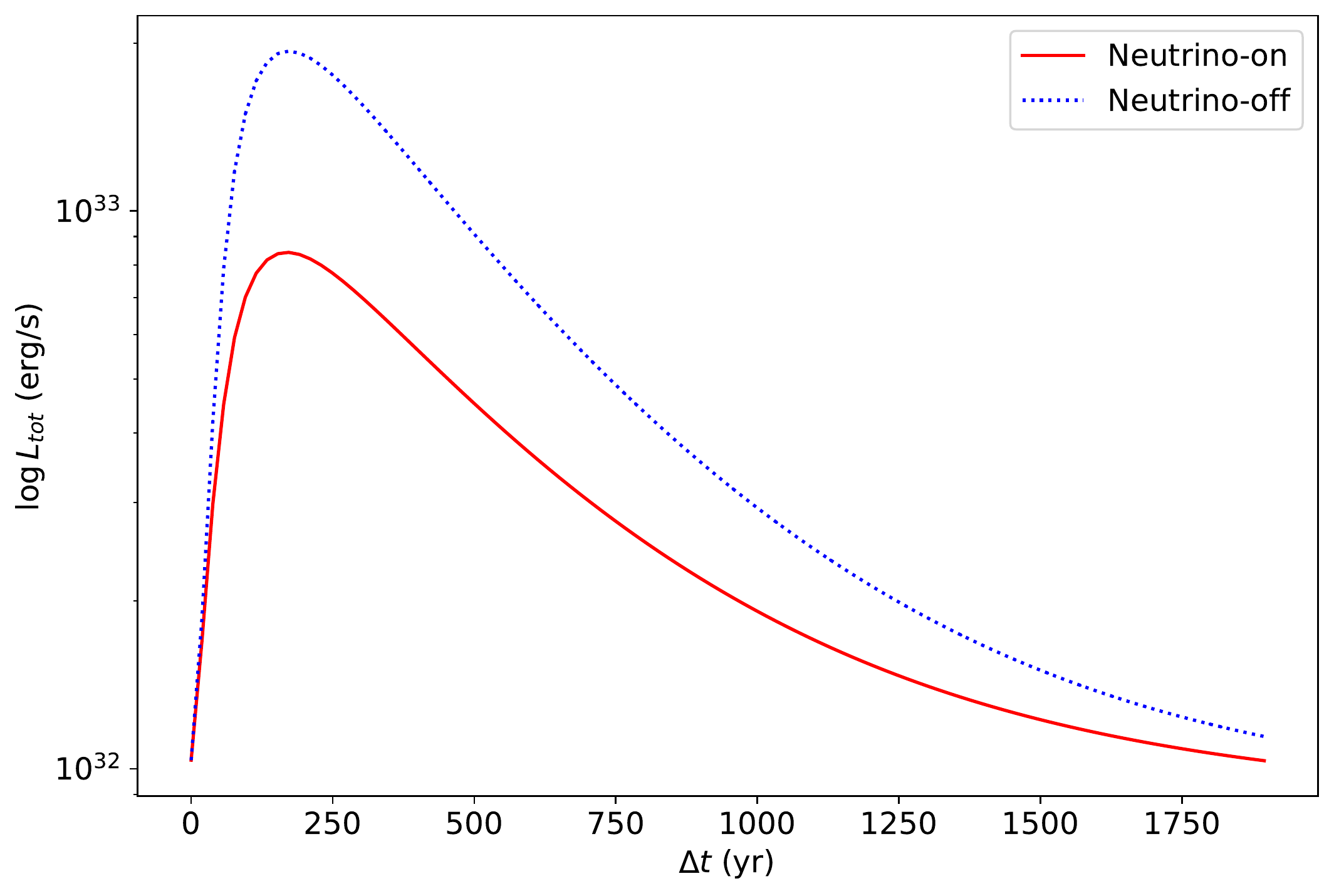}
    \caption{Luminosity evolution after an impulsive heat injection in the inner half of the crust. Neutrino emission reduces the peak luminosity by a factor $\sim 2.3$ compared to radiative cooling only.}
    \label{fig:decay}
\end{figure}

In our test model heat has been released in the northern hemisphere and in the innermost half of the crust, assuming a gaussian profile along the three spatial dimensions with $\sigma_r\simeq\SI{100}{\meter}$, $\sigma_\theta\simeq\sigma_\phi\simeq\SI{\pi/5}{\radian}$. The additional heating term in equation (\ref{eq:T}) is $\dot H \simeq\SI{5e37}{\erg\per\second}$, modulated by the gaussian profile. Heating is assumed to be quasi-instantaneous (the $\dot H$ term is active for $\Delta t_\text{inj}\simeq\SI{3}{\second}$). The NS luminosity has then been calculated assuming blackbody emission at the local temperature, after deriving $T_s$ from equation (\ref{eq:tsurf}). The time evolution follows a typical FRED (fast-raise-exponential-decay) pattern, as shown in Fig. \ref{fig:decay}. The two curves in Fig. \ref{fig:decay} illustrate the role played by neutrino losses in the crust. The temperature rise produced by heat deposition, in fact, is large enough in this case to make neutrino emission sizable (contrary to what occurs when the crust is not heated) and this results in a photon luminosity lower by a factor $\sim 2$ with respect to the case in which neutrino losses are turned off. In the present case the peak luminosity is $\sim \SI{e33}{\erg\per\second}$, with an increase of a factor $\approx 10$ above the quiescent level.

\begin{figure}
    \centering
    \includegraphics[width=.48\textwidth]{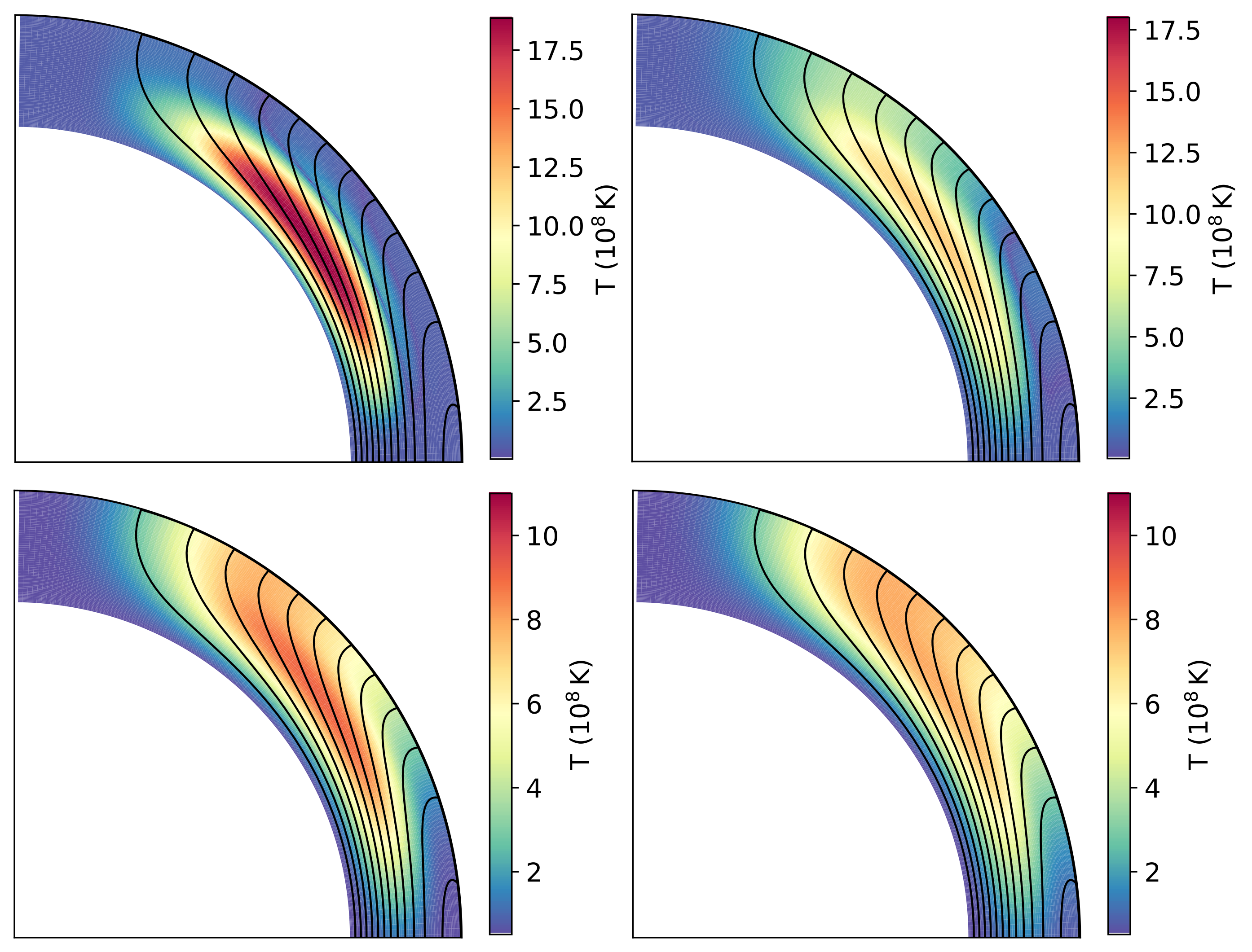}
    \caption{Meridional cuts (at the same $\phi$) of the evolution of the hot spot during the rise phase. \added{The first panel corresponds to the initial injection, and the subsequent ones are separated by $\sim\SI{50}{\year}$.} Transport of heat to the surface happens preferentially along magnetic field lines, whose planar projection is superimposed in black. Note that colour bar range decreases between the two rows to improve visualisation.}
    \label{fig:drift_cut}
\end{figure}

\begin{figure}
    \centering
    \subfigure{\includegraphics[width=.23\textwidth, trim=180 160 80 150,clip]{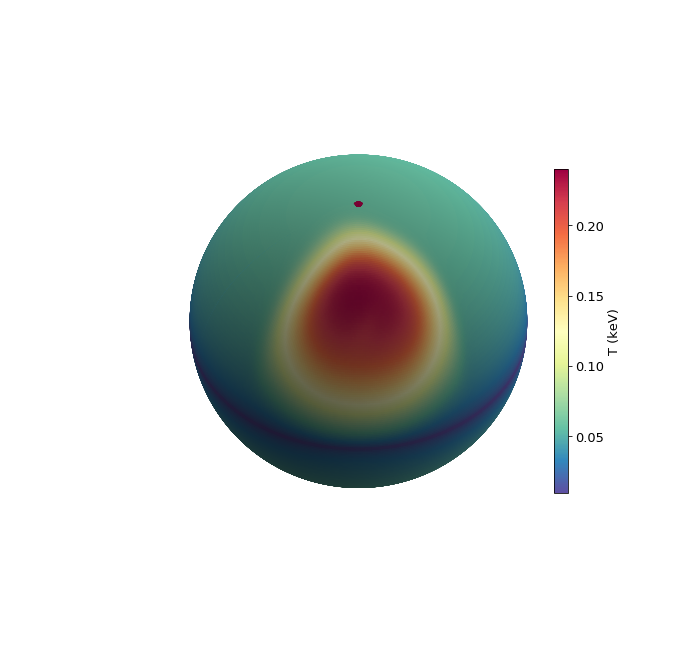}}
    \subfigure{\includegraphics[width=.23\textwidth, trim=180 160 80 150,clip]{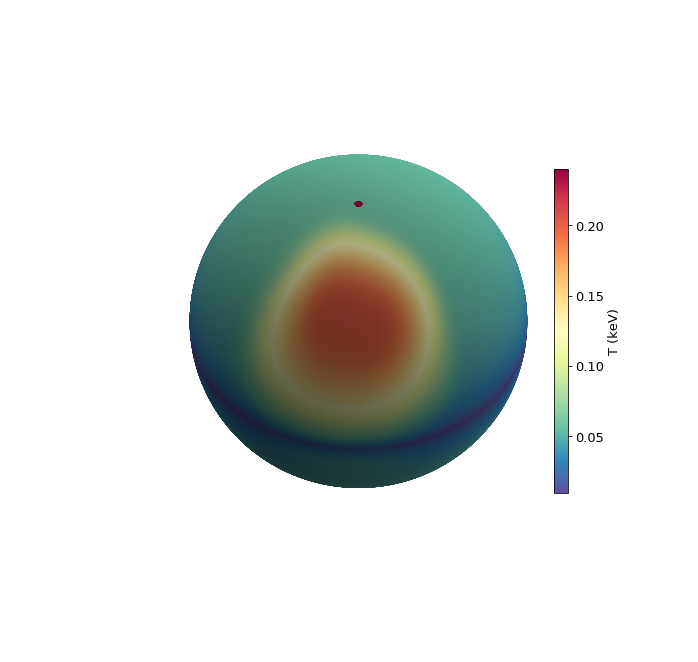}}  
    \subfigure{\includegraphics[width=.23\textwidth, trim=180 160 80 150,clip]{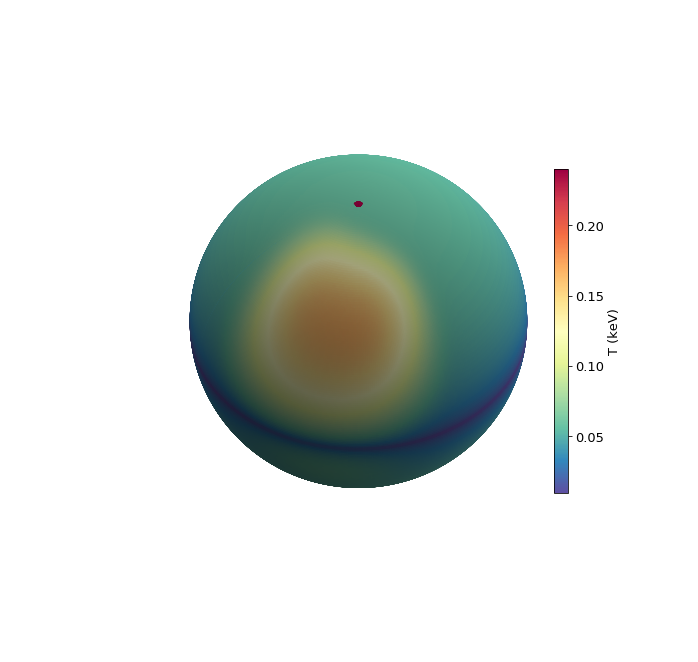}}  
    \subfigure{\includegraphics[width=.23\textwidth, trim=180 160 80 150,clip]{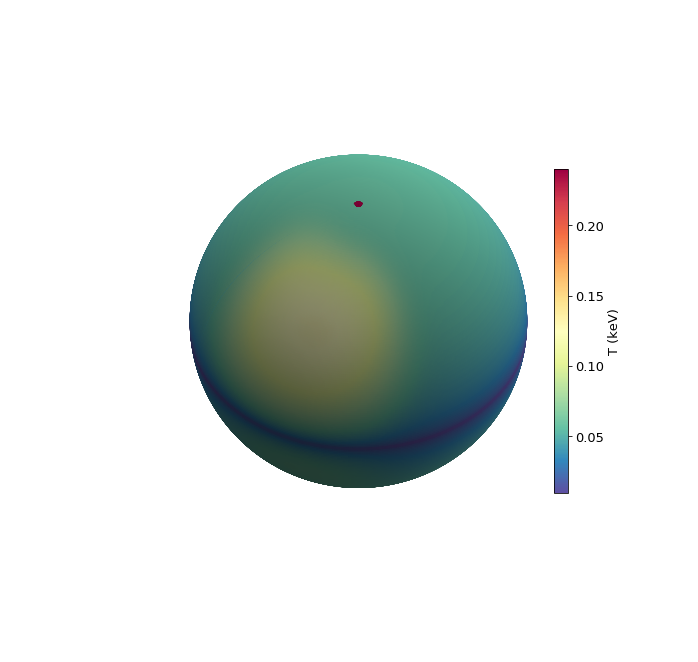}}
    \caption{Surface thermal evolution of the hot spot producing the luminosity shown in Fig. \ref{fig:decay}. Time increases from left to right and from top to bottom; snapshots are separated by $\sim\SI{200}{\year}$ and the first one corresponds to the peak of the luminosity curve. The magnetic north pole is highlighted for reference.}
    \label{fig:drift_spot}
\end{figure}

The hot structure that develops onto the surface exhibits a somehow peculiar evolution. In fact, its shape is determined by heat diffusion, which is not isotropic but depends on the magnetic field direction according to equations (\ref{eq:sigma}) and (\ref{eq:wf}).
Hence, heat tends to flow along field lines. Figure \ref{fig:drift_cut} shows how during the luminosity rise time heat is not just flowing radially to reach the surface, but does so following the magnetic field.
Moreover, once it is formed the hotter region tends to drift as it cools down, both in latitude, towards the equator, and in longitude. Such behaviour is clearly visible in Figure \ref{fig:drift_spot} which shows four snapshots of the heated surface patch evolution.

The duration of this event is of some thousands of years, with a rise time of about a century (however, timescales in this case are affected by code limitations, see section \ref{sec:limits}). Still, on a qualitative basis, it can be taken as a representation of the observed flux variation during  magnetar outbursts, which happens on shorter timescales \citep{2018MNRAS.474..961C}. 

\subsection{Surface heating}
\label{sec:surfheating}
The framework discussed in Section \ref{sec:deepheating} can be easily adapted to study the shape of pulsar hot spots. In fact, the physical ingredients remain the same as long as it can be assumed that no other effects apart from heating come into play. The major difference with respect to the model presented in the previous section is that now energy is deposited in the outermost crustal layers, as it is the case e.g. for the heat deposited by backflowing currents on the surface of radio-pulsars. 

\begin{figure}
    \centering
    \includegraphics[width=.4\textwidth]{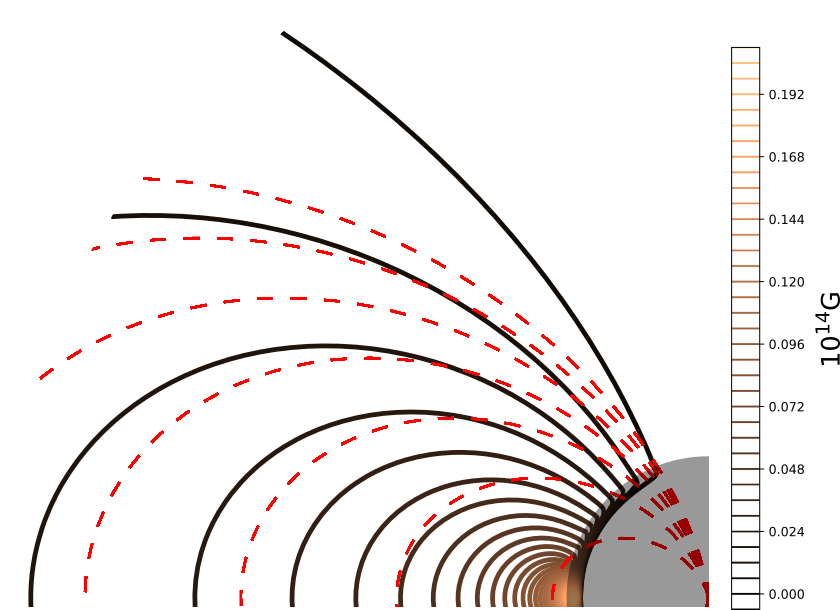}
    \caption{The extrapolated external magnetic field lines, in which the colour bar encodes the strength from black (zero) to copper (maximum value), compared to a purely dipolar field (dashed red lines). Given the high degree of symmetry of this case, only a quarter of the star is shown for better visualisation.}
    \label{fig:magsphere}
\end{figure}

As a background state, we take a NS with initial field $B_{pol}\approx B_{tor}\simeq\SI{e12}{\gauss}$ and temperature $T\simeq\SI{5e7}{\kelvin}$, which was evolved for some Hall times, $t\sim\SI{e5}{\year}$. Then, a heating source is placed in a small region of size $\sim\SI{0.5}{\kilo\meter}$ close to the magnetic pole. We chose this spot position since even though the external magnetic field in our simulations is not a pure dipole, its qualitative shape is similar to it, as displayed in figure \ref{fig:magsphere}, and heating of the polar regions is to be expected.\footnote{Even if our simulations do not explicitly include the dynamics of the magnetosphere, its configuration can be extrapolated from the boundary condition \ref{BC_Bext} for the magnetic field at the top of the crust, requiring that for each harmonic $B_\ell^m(r>R_\star)=B_\ell^m(R_\star)/r^{\ell+1}$.\label{foot:BC}} 

Backflowing currents in pulsars can reach a depth ranging from about a tenth to the entire width of the crust \citep{10.1093/mnras/stz1507}.
In our simulations we choose to insert the heat source uniformly from the surface down to a quarter of the crust width. We at any rate checked that different depth values provide quite similar results, possibly because this length is anyway much smaller than the other relevant lengths in the problem.

\begin{figure}
\centering
\subfigure{
   \includegraphics[width=.23\textwidth,trim=120 120 65 90, clip]{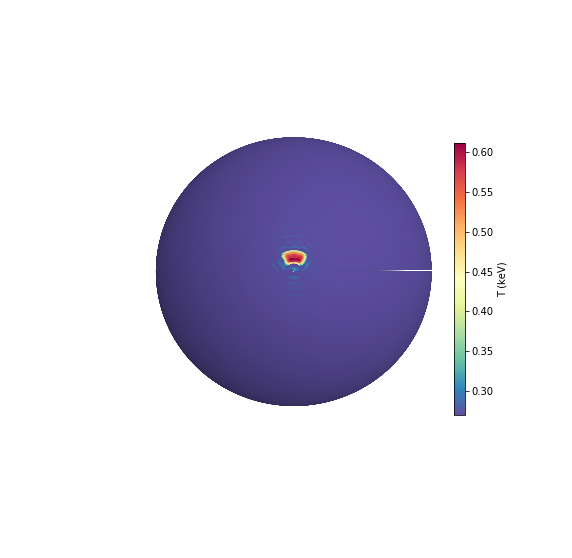}}\!%
\subfigure{
  \includegraphics[width=.23\textwidth,trim=120 120 70 90, clip]{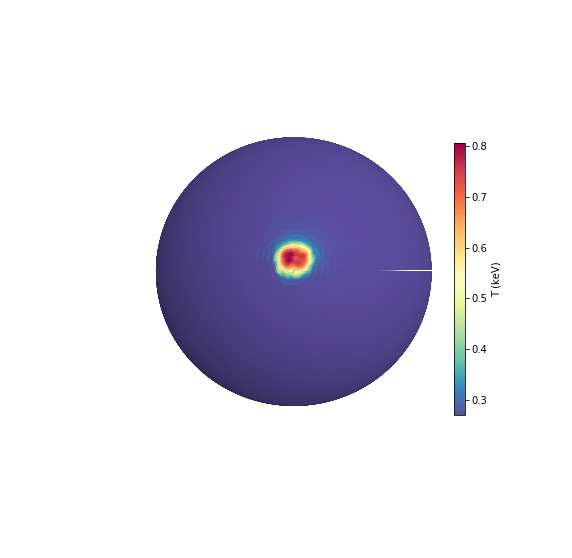}}\\
\subfigure{
   \includegraphics[width=.23\textwidth,trim=120 90 50 90, clip]{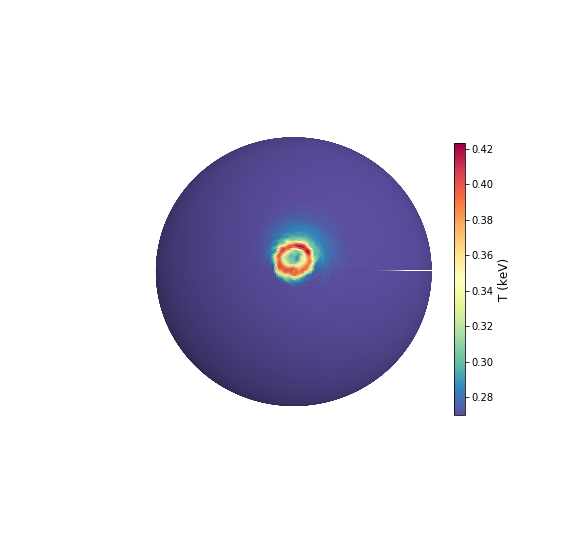}}\!%
\subfigure{
  \includegraphics[width=.23\textwidth,trim=120 90 50 90, clip]{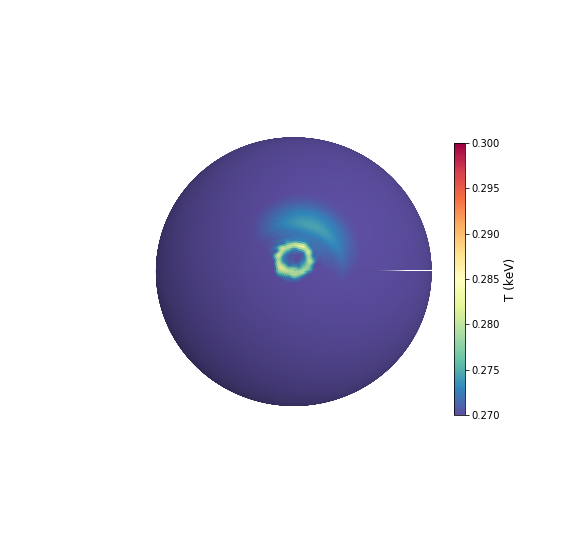}}
  \caption{\emph{Top row:} initial (left) and equilibrium (right) stages of the evolution of a hot region (magnetic pole at the centre, towards the observer) during steady heating from above. \emph{Bottom row:} cooling of the spot after heating is turned off. The last three snapshots (from top to bottom and from left to right) are separated by a time interval $\sim\SI{200}{\year}$. Note that the colour bar range decreases between panels to highlight the effect.}
\label{spot_onset}
\end{figure}

Results show that if heat injection is steady, the hot spot reaches a state of quasi-equilibrium in a few years. Starting with a heated patch of size $\sim\SI{1}{\kilo\meter}$, the spot tends to assume a quasi-circular shape, staying on top of the injection region. The evolution is shown in figure \ref{spot_onset} (top row), where the initial shape and the equilibrium configuration of the hot spot are compared for a steady heat injection $\dot H =\SI{5e25}{\erg\per\second}$.
We then followed the evolution of the same spot after the heating term is turned off. In a time $\approx\SI{1000}{\year}$, the spot cools down in such a way that only a ring corresponding to the region rim is left. In the final cooling phases (\ref{spot_onset}, bottom row), a crescent like structure drifting towards the equator becomes visible. It is a somehow subdominant feature, since the main hot spot is still in correspondence to the initially heated region but we observed that it can eventually become hotter than the central spot in the very late cooling stages (when temperature differences from the background are small). \added{We note that some asymmetry in the position/shape of the initially heated region with respect to the magnetic pole is necessary for the formation of crescent-like structures during the evolution. Heat injection in a circular patch exactly below the magnetic pole results in a nearly circular cooling spot. However, simulations show that such deviations need to be indeed small and in real sources they are expected to be produced, e.g.,  by the effect of the coupling of crustal heating currents with the rotation of the star \citep{10.1093/mnras/stz1507} or the presence of sub-dominant non-dipolar components.}

\begin{figure}
\centering
\subfigure{
\includegraphics[width=.22\textwidth,trim=120 90 60 90, clip]{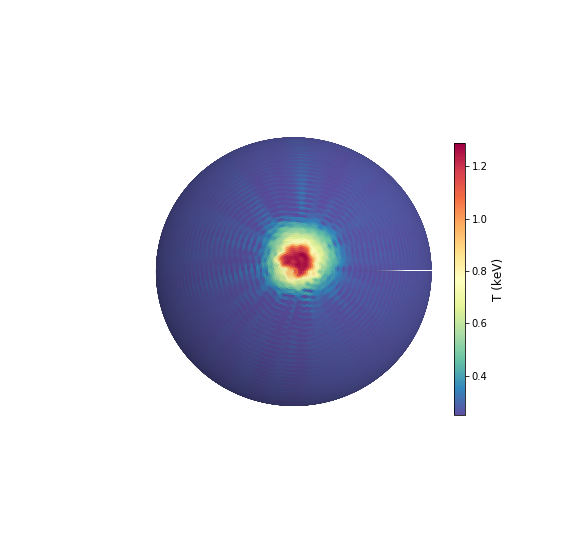}}\!
\subfigure{
    \includegraphics[width=.22\textwidth,trim=120 90 60 90, clip]{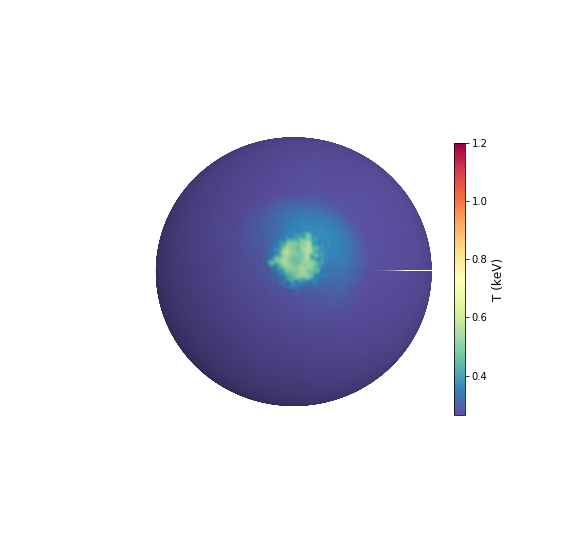}}\!
\subfigure{
  \includegraphics[width=.22\textwidth,trim=120 90 60 90, clip]{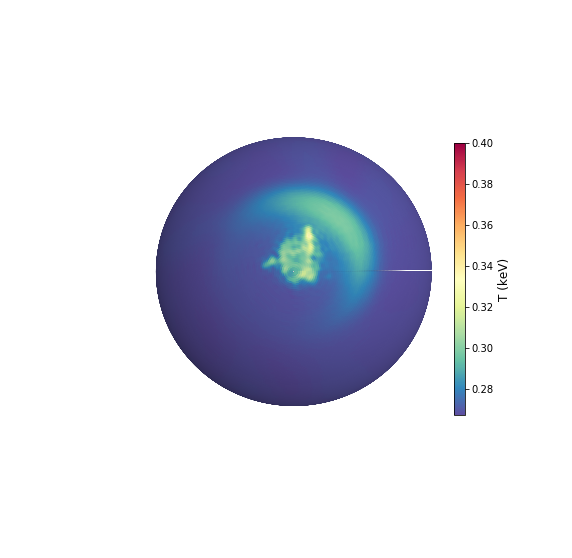}}\!
\subfigure{
  \includegraphics[width=.22\textwidth,trim=120 90 60 90, clip]{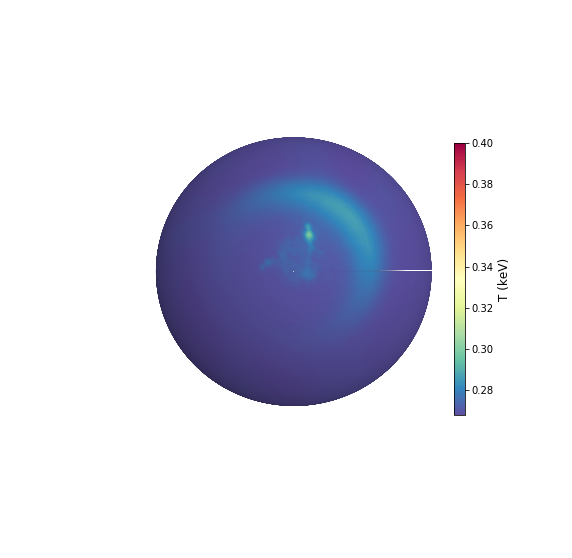}}
\caption{Cooling of a spot similar to the one in Figure \ref{spot_onset} but for a heat flux 10 times higher. Here snapshots are separated by $\sim\SI{300}{\year}$, the first one referring to the time at which heating stops. Note that colour bar range decreases between the two rows to highlight the effect.}
\label{spot2}
\end{figure}

In the case of a higher heat injection, $\dot H= \SI{5e26}{\erg\per\second}$, we observe a similar phenomenology, but a turbulent like pattern emerges, see fig. \ref{spot2}, \added{which is nonetheless well resolved by our grid}.
This results in a more complex evolution of the shape as the spot cools down. Its relic, in fact, gets fragmented into many smaller structures that do not exhibit the ordered, ring-like shape of the previous case. A drifting crescent-shaped subdominant structure is again formed in the final phases.

\begin{figure}
\centering
\subfigure{
  \includegraphics[width=.23\textwidth]{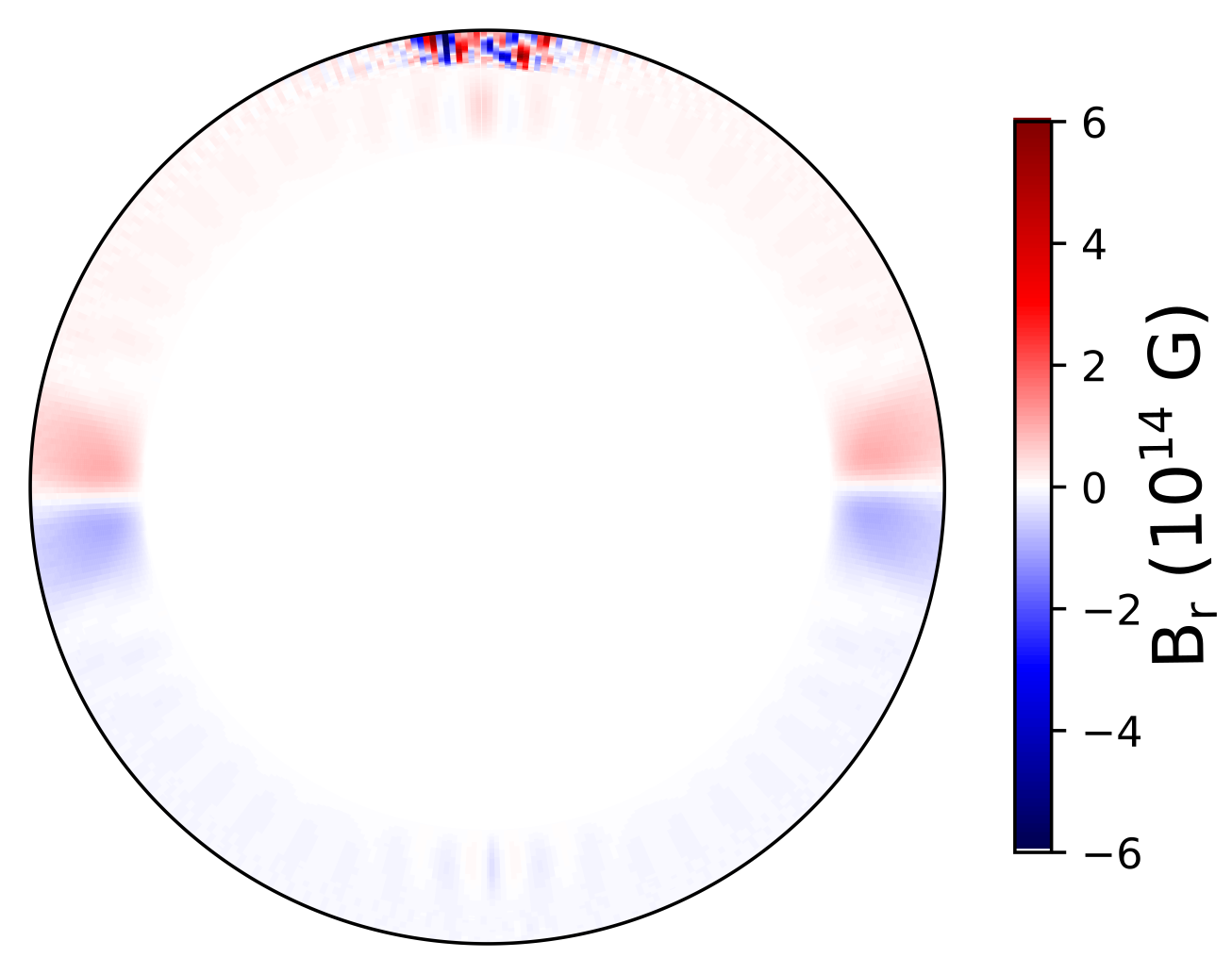}}
  \subfigure{
  \includegraphics[width=.22\textwidth]{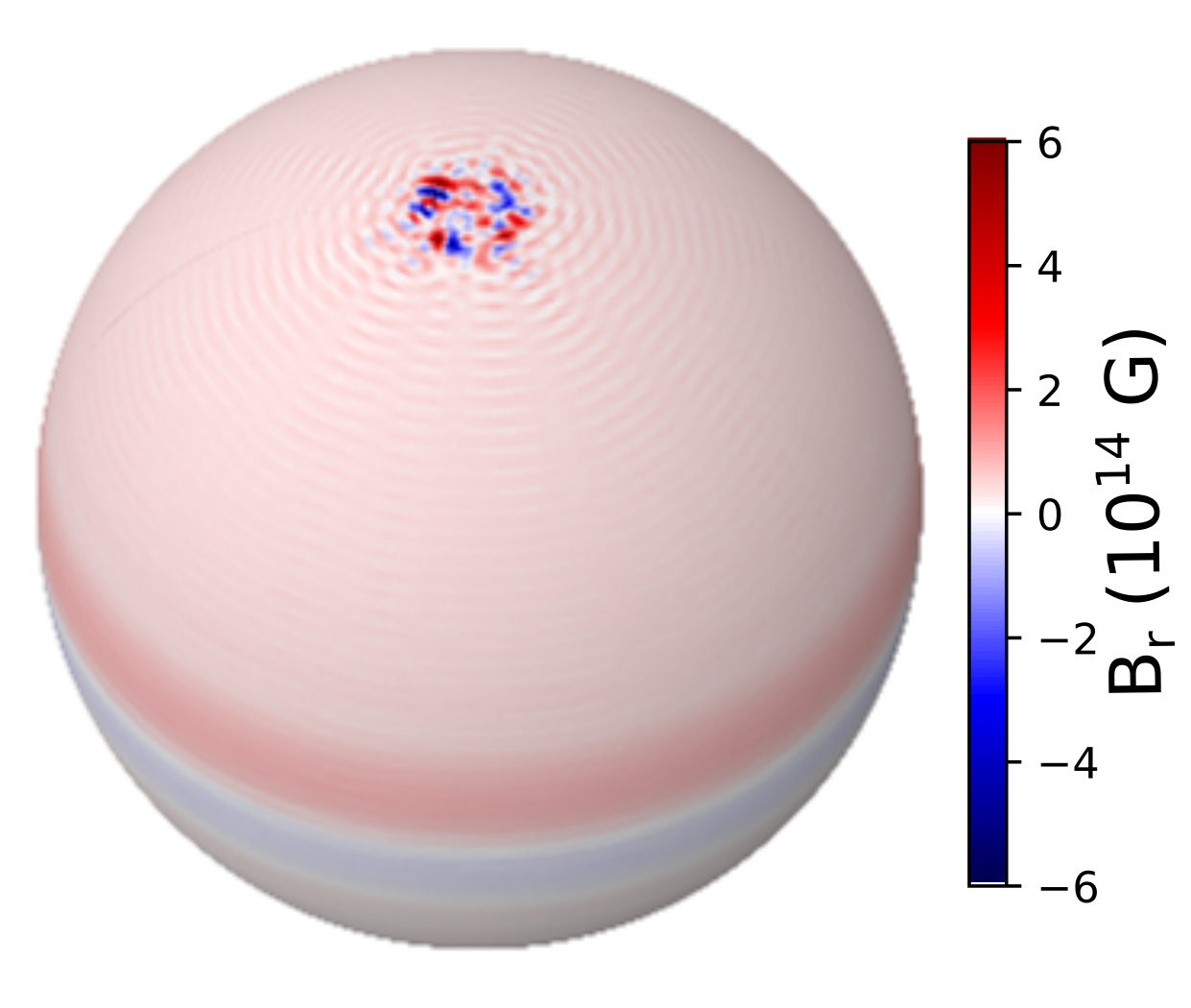}}
    \caption{Radial component of the magnetic field in the localised heating steady state shown on the first panel of figure \ref{spot2}. The crust thickness is enhanced by a factor $4$ to help visualisation.}\label{fig:Benh}
\end{figure}

Moreover, the backreaction on the magnetic field become important: in fact, in this situation a temperature gradient perpendicular to the (radial) density one develops, hence the Biermann battery effect, that provides a negligible feedback in the long term evolution of isolated NSs, can give rise to a substantial local enhancement of the magnetic field. Such behaviour is displayed in figure \ref{fig:Benh}. When the stationary state is reached, some small magnetic structures appears on top of the $\approx\SI{e12}{\gauss}$ large-scale (quasi) dipolar field, where the field strength can reach values up to $\SI{6e14}{\gauss}$. Thus, localised heating may also account for small-scale magnetic structures in the crust, that are therefore not originated by dynamo-like processes.

\begin{figure}
    \centering
    \includegraphics[width=.48\textwidth]{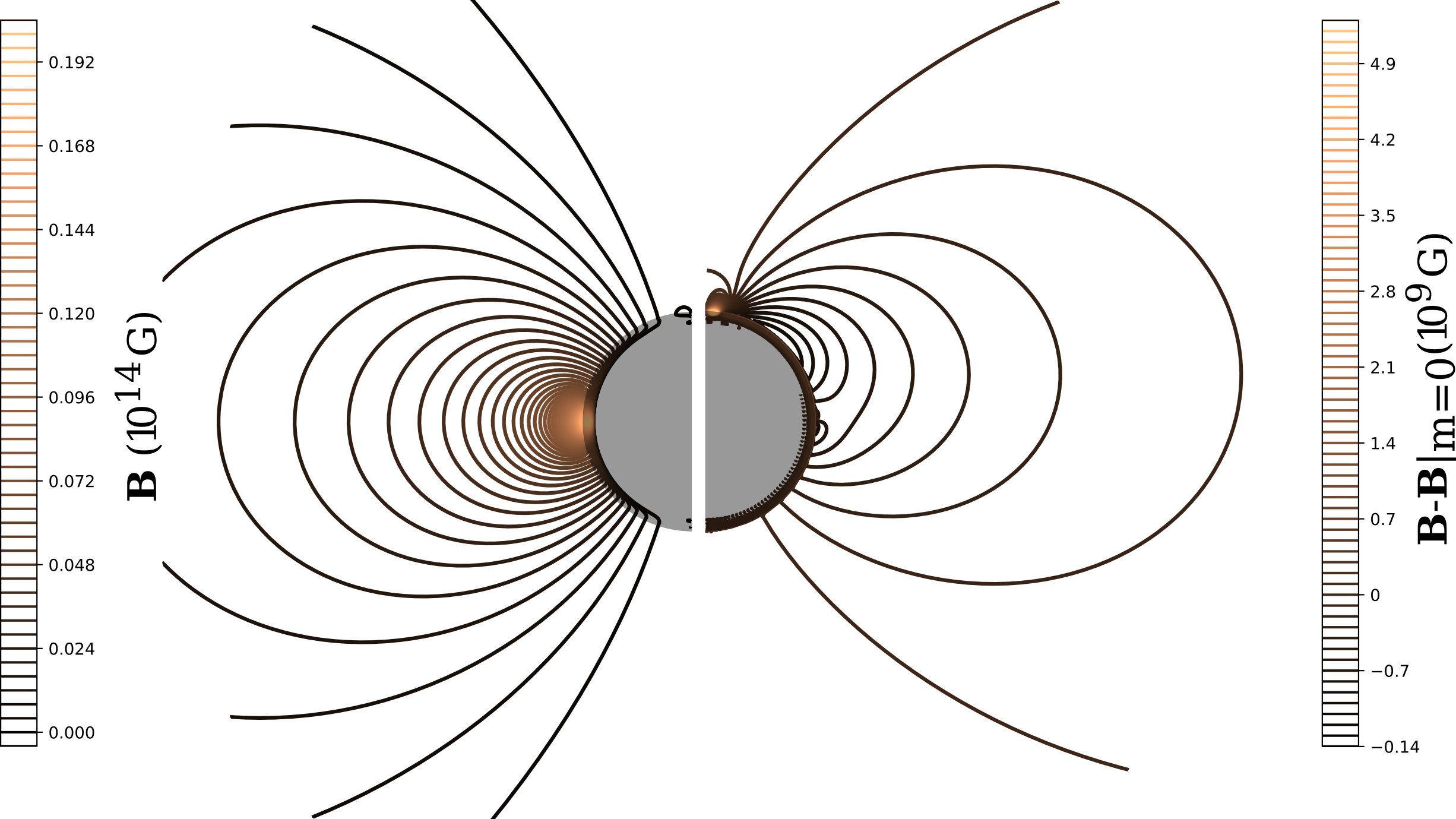}
    \caption{The extrapolated external magnetic field for the case in Figure \ref{fig:Benh}. The left half shows the total field and the right one the difference between the total field and its $m=0$ modes, which are dominated by the dipole.}
    \label{fig:magnsphere_2}
\end{figure}

\added{The appearance of these crustal magnetic features reflects in the creation of local magnetic structures in the magnetosphere, even though the overall B-field remains very close to dipolar. Figure \ref{fig:magnsphere_2} shows the external field lines for the same case as in Figure \ref{fig:Benh},  as derived by solving the magnetospheric structure (see  footnote \ref{foot:BC} at page \pageref{foot:BC}). After subtracting the contribution of the $m=0$ modes, which are dominated by the dipolar field, a small magnetic field loop is clearly visible above the heated region, extending outwards up to a distance $\lesssim R_\star$ with a typical strength $\approx\SI{e9}{\gauss}$. This shows that magnetic structures are not necessarily confined to the crust, but can extend in the inner magnetosphere.
}

The cooling phase lasts some thousands years. It is therefore possible that the aftermath of powerful heating events can produce long-lasting thermal structure on a NS crust, evolving in complex patterns along field lines. 

\section{Discussion and conclusions}\label{sec:conclusions}
In this paper we presented for the first time 3D numerical simulations of the coupled magneto-thermal evolution in isolated neutron stars with full account for neutrino emission from the crust and a simplified neutrino core cooling model. While results for the long-term evolution show no substantial deviations with respect to those obtained with 1D and 2D calculations \cite[see e.g.][for a review]{Pons_2019}, the capability of a 3D approach to consistently deal  also with the smaller spatial scales proved essential to highlight the onset of eMHD instabilities and to follow the evolution of localised heat injection in the star crust. In particular, out main findings are:
\begin{enumerate}[label=(\roman*)]
    \item the magnetic field evolves towards the so called Hall attractor \citep{2014PhRvL.112q1101G}. In this configuration magnetic energy is stored \added{preferentially in the odd modes and especially in the $\ell=1,2,3,5$ ones}. This results in the appearance of magnetic and thermal structures near the (magnetic) equator;
    \item a strong toroidal magnetic field component ($\approx$\SIrange[range-phrase = --]{e15}{e16}{\gauss}) can trigger the resistive tearing eMHD instability in less than a Hall time. Our simulations show that the appearance of small-scale high-B structures, mainly along the equator, coupled with a local enhancement of the temperature produce the conditions for crustal yielding according to the von Mises criterion;
    \item a localized, impulsive heat injection in the deep crustal layers results in a cooling hot spot on the star surface. The emitted luminosity has a sharp rise followed by an longer decay; 
    \item as a result of anisotropic heat transport in the magnetized crust, the heated region drifts and may change its shape as it cools; 
    \item even with an essentially dipolar field, quasi symmetric hot regions near the poles can cool down assuming a crescent-like shape.
\end{enumerate}

%\paragraph{Localised heating}
Our 3D simulations of the evolution of a locally heated region in the star crust revealed a variety of behaviours reflecting the location of the heat source (position and depth in the crust), the energy injection rate and the crust magnetic and thermal structure. In particular we considered two scenarios, in which heating occurs either inside the crust (deep heating) or in the outermost layers (surface heating). Both of them may be relevant for magnetar outbursts, during which a hotter region on the star surface appears and then progressively cools down and shrinks \cite[see e.g.][]{2018MNRAS.474..961C}. In fact, this has been explained in terms of dissipated magnetic energy inside the crust \citep{2002ApJ...580L..69L,2012ApJ...750L...6P} or of Joule heating due to returning currents flowing along the field lines of a (locally) twisted magnetic field \citep[][see also \citealt{2015RPPh...78k6901T}]{2009ApJ...703.1044B}.

%\paragraph{Neutrino emission}
In the simulations presented in this work, neutrino emission is relevant only for the case presented in section \ref{sec:deepheating}. In fact, neutrinos become important if injection is fast, so that high local temperatures can be reached. According to 2D simulations \cite{2012ApJ...750L...6P}, large neutrino losses result in an  upper limit on the radiative luminosity released in magnetar outbursts. At present such regimes can not be investigated with our code, due to numerical hindrances associated with the treatment in three dimensions of a strongly non linear term (as a rule of thumb, $N_\nu\propto T^{7.5}$).
In fact, this term can cause the appearance of numerical spurious features in our solutions when temperature gradients become very high. Moreover, if the background state has an ultra-strong magnetic field, turbulent patterns analogous to those discussed in section \ref{sec:surfheating} can be triggered also in a context of impulsive heat injection; with our present numerical set-up, such behaviour has proven to be hard to treat numerically when neutrino losses are important. This prevents a comprehensive treatment of impulsive heating events. The question if (and how) results obtained in a 3D framework are different with respect to the 2D treatment of \cite{2012ApJ...750L...6P} is a matter that will be addressed in a future study.

\subsection{Ramifications}\label{sec:discussion}

\paragraph{Comparison with low-B magnetars} The presence of strong toroidal fields in magnetars has been invoked since long to explain their distinguishing activity as compared to radio-pulsars with similar spin-down magnetic fields, the high-B pulsars with $B_\text{dip}\approx \num{e13}$--$\SI{e14}{\gauss}$ \cite[see e.g.][]{2015RPPh...78k6901T}. On the other hand, some sources with $B_\text{dip}$ as low as $\approx \SI{e12}{\gauss}$ can show magnetar-like activity \cite[the low-B magnetars; see e.g.][and references therein]{2011ApJ...740..105T}. According to our simulations, the resistive tearing instability appears on a timescale $\lesssim \tau_H\approx 10^4$ yr and lasts for about $\approx 1000$ yr. This mechanism can hence provide a viable explanation for the activity (bursts and outbursts) detected in young sources (age $\lesssim 10^4$ yr), which are the vast majority of the magnetar population.\footnote{See the McGill magnetar catalogue at {\url{http://www.physics.mcgill.ca/~pulsar/magnetar/main.html}} \citep{2014ApJS..212....6O}.} Whether such an instability can be triggered under the conditions typical of older objects, like the low-B sources {SGR~0418+5729} and {Swift~J1822.3-1606} \cite[age $\approx 10^5$--$10^6$ yr;][]{2011ApJ...740..105T, 2012ApJ...754...27R} or the onset of outbursts is produced by a different, possibly related, mechanism is an open question.

\paragraph{Crescent-shaped features and observations} Non-polar, crescent-like hot spots have been recently detected in NICER X-ray observations of the millisecond pulsar {PSR~J0030+0451} and interpreted as due to heating from backflowing currents in a non-dipolar magnetic field \citep{miller}. Our results show that such features can actually form as thermal relics of past events of heat deposition even in presence of a dipole-dominated field, provided that proper account for the crustal transport properties is made. Even though the evolutionary history of {PSR~J0030+0451} is likely quite different from that of a passively cooling NS, and its (dipolar) field is lower than the one used in our model, our results show that a qualitatively similar behaviour of the crust may be responsible of the observed pattern even without invoking strong multipolar field components.

\paragraph{Battery effects and magnetar magnetospheres} In section \ref{sec:heating} we showed how the magnetic field created through battery effects by an external heating source can reach strong local values in a turbulent-like pattern. \added{The existence of small-scale magnetic structures, in which the field strength is orders of magnitude higher than in the surrounding dipole, has been invoked to explain the (relatively) large energy ($\approx$ \SIrange[range-units=single, range-phrase = --]{1}{10}{\kilo\electronvolt}) of absorption features detected in the (quiescent) emission of some magnetars, if these are interpreted as due to cyclotron absorption/scattering onto protons, $E_{\text{cp}}\simeq 0.6\,(B/\SI{e14}{\gauss})$ keV. The prototypical source is the low-field ($B_{\text{dip}}\sim \SI{6e12}{\gauss}$) magnetar {SGR~0418+5729}, where a phase-dependent absorption feature at $\sim$\SIrange[range-units=single, range-phrase = --]{2}{10}{\kilo\electronvolt} was discovered in the {\em XMM-Newton} data by  \cite{2013Natur.500..312T}. According to their interpretation, the line arises as radiation from a cooling spot on the star surface crosses a baryon-loaded, small ($\approx \SI{100}{\meter}$) magnetic loop where absorption occurs. Albeit associating this kind of magnetic structure with those produced  by the battery effect in our simulations is tempting, we warn that thermocoupling effects turn out to be less important in the case of deep heating (see section \ref{sec:deepheating}), where the local enhancement of the magnetic field is modest.}

\subsection{Present limitations}\label{sec:limits}
In this work, we highlighted the perspectives that a novel three dimensional approach can open in the study of neutron star magneto-thermal evolution. There are, nevertheless, some limitations that must be taken into account when interpreting our numerical results.

In fact, we had to reduce the microphysical input to a realistic yet simplified model for the computing time to be manageable. This concerns in particular the use of a simplified form for hydrostatic equilibrium density profile of equation  (\ref{n_profile}), which was also assumed independent on temperature and magnetic field, and the use of a constant $\tau$ throughout the crust.
Moreover, we have chosen some strong prescriptions on thermal conductivity and heat capacity. In particular, the assumption that $C_V$ is linearly dependent on the temperature is valid only for the electron contribution, and does not take into account the contribution of the lattice. This implies that equation (\ref{eq:heat}) depends on $T^2$ only, which is a key point for the efficiency of the numerical scheme. However, such an assumption becomes questionable when the term $\propto\partial T/\partial t$ starts to dominate, as in the case of impulsive heating in section \ref{sec:deepheating}. \added{In particular, this affects our estimates for the duration of thermal relaxation events. In fact, the heat diffusion timescale across a length $L$ can be estimated as \citep{2018Ap&SS.363..209C}
\begin{equation}
    \tau_{\text{diff}}\sim\frac{1}{4}\left[\int_L \text{d}l \left(\frac{C_V}{\kappa}\right)^{1/2}\right]^{2},
\end{equation} hence it is regulated by the specific heat-to-thermal conductivity ratio. According to the estimates of \citet{2018Ap&SS.363..209C}, for the typical conditions of a NS, the timescale of heat transport from an internal heater to the surface is $\lesssim\SI{1}{\year}$, whereas in our model the value of the characteristic diffusion time across the crust turns out to be much longer, $\tau_\text{diff}\approx\SI{50}{\year}$. This may well be related to our assumptions which make the ratio $C_V/\kappa$ is independent on temperature, while it is expected to depend from temperature as well as from the properties of crustal superfluidity \citep[][and references therein]{transport}. Hence, our results for this case should simply be regarded as indicative of the general evolution of such events. For the model discussed above the evolution timescale is longer than what expected under more realistic conditions by a factor $\approx 100$, although extending this to other cases is haphazard.}

Another strong prescription is that the whole physics of the core is embodied in the boundary conditions (\ref{eq:core-cool}) and (\ref{eq:typeI}). Addressing the complex microphysics of the core and the description of the crust-core transition is beyond the scope of this paper (and is in general a problem best suited for one-dimensional studies). However, a direct implication of equation (\ref{eq:core-cool}) is that heat conduction from the crust to the core is inhibited. This is not a problem for our models, but could become an issue when dealing with extremely high heat injections in the deep crust.
Moreover, equation (\ref{eq:typeI}) implies that the core is assumed to be in a Type I superconducting phase, so that no magnetic field is allowed to enter it, \added{and that the magnetic flux has been completely expelled during the phase transition}. However, current models of pulsar glitches \citep[see e.g.][]{1969Natur.224..872B} suggest that the state is of a Type II superconductor, \added{or in any case that at least} some field is present in the core. Dealing with the modelling of this more complicated transition is again beyond the scope of this work.

\acknowledgments
Simulations were run at CloudVeneto, a HPC facility jointly owned by the University of Padova and INFN, and at UCL Grace HPC facility (Grace@UCL). The authors gratefully acknowledge the use of both facilities and the associated support services. RT and RT acknowledge financial support from the Italian MIUR through grant PRIN 2017LJ39LM. 
\vspace{5mm}

\bibliography{biblio}{}
\bibliographystyle{aasjournal}



\end{document}